\PassOptionsToPackage{table,xcdraw}{xcolor}
\documentclass[acmsmall]{acmart}
\usepackage{pdflscape}
\usepackage{comment}
\usepackage{xcolor}
\usepackage{float}
\usepackage{balance}
\usepackage{xspace}
\usepackage{subcaption}
\usepackage{multirow}
\usepackage{placeins}
\usepackage{placeins}
\usepackage{cleveref}
\hypersetup{citecolor = blue}
\usepackage[framemethod=TikZ]{mdframed}
\usepackage[utf8]{inputenc}
\graphicspath{{figures/}}
\usepackage[font=small,format=plain,labelfont=bf,textfont=it]{caption}
\usepackage{booktabs}\usepackage[all=normal,wordspacing]{savetrees}
\usepackage{enumitem}

\crefformat{section}{\S#2\color{blue}#1#3} 
\crefformat{subsection}{\S#2\color{blue}#1#3}
\crefformat{subsubsection}{\S#2#1#3}

\pgfsetarrows{latex-latex}
\tikzset{%
  base/.style = {inner sep=5pt,
                 text centered,
                 thin,
                 font=\rmfamily},
  round/.style = {base,
                  rectangle,
                  rounded corners=1ex,
                  draw=black,
                  fill=gray!20,
                  minimum height=0.35in}
}

\newcommand{\ookla}{Ookla\xspace}
\newcommand{\ndt}{NDT7\xspace}

\renewcommand{\paragraph}[1]{\vspace*{0.015in}\noindent\textbf{#1}}
\renewcommand{\shortauthors}{Kyle MacMillan et al.}

\AtBeginDocument{%
  \providecommand\BibTeX{{%
    \normalfont B\kern-0.5em{\scshape i\kern-0.25em b}\kern-0.8em\TeX}}}

\setcopyright{acmlicensed}
\acmJournal{POMACS}
\acmYear{2023} \acmVolume{7} \acmNumber{1} \acmArticle{19} \acmMonth{3} \acmPrice{15.00}\acmDOI{10.1145/3579448}

\received{October 2022}
\received[revised]{December 2022}
\received[accepted]{January 2023}



\begin{document}
\setlist[itemize]{leftmargin=*}	

\title{A Comparative Analysis of Ookla Speedtest and Measurement Labs Network
Diagnostic Test (NDT7)}


\author{Kyle MacMillan}
\affiliation{%
  \institution{University of Chicago}
  \city{Chicago}
  \country{USA}}
\email{macmillan@uchicago.edu}

\author{Tarun Mangla}
\affiliation{%
  \institution{University of Chicago}
  \city{Chicago}
  \country{USA}}
\email{tmangla@uchicago.edu}

\author{James Saxon}
\affiliation{%
  \institution{University of Chicago}
  \city{Chicago}
  \country{USA}}
\email{jsaxon@uchicago.edu}

\author{Nicole P. Marwell}
\affiliation{%
  \institution{University of Chicago}
  \city{Chicago}
  \country{USA}}
\email{nmarwell@uchicago.edu}

\author{Nick Feamster}
\affiliation{%
  \institution{University of Chicago}
  \city{Chicago}
  \country{USA}}
\email{feamster@uchicago.edu}

\renewcommand{\shortauthors}{Kyle MacMillan et al.}

\begin{abstract}
	Consumers, regulators, and ISPs all use client-based ``speed tests'' to
	measure network performance, both in single-user settings and in aggregate.
	Two prevalent speed tests, Ookla's Speedtest and Measurement Lab's Network
	Diagnostic Test (NDT), are often used for similar purposes, despite having
	significant differences in both the test design and implementation, and in
	the infrastructure used to perform measurements.  In this paper, we present
	the first-ever comparative evaluation of \ookla and \ndt (the latest version
	of NDT), both in controlled and wide-area settings.  Our goal is to
	characterize when and to what extent these two speed tests yield different
	results, as well as the factors that contribute to the differences. To study
	the effects of the test design, we conduct a series of controlled, in-lab
	experiments under a comprehensive set of network conditions and usage modes
	(e.g., TCP congestion control, native vs. browser client). Our results show
	that \ookla and \ndt report similar speeds under most in-lab conditions,
	with the exception of networks that experience high latency, where \ookla
	consistently reports higher throughput. To characterize the behavior of
	these tools in wide-area deployment, we collect more than 80,000 pairs of
	\ookla and \ndt measurements across nine months and 126 households, with a
	range of ISPs and speed tiers. This first-of-its-kind paired-test analysis
	reveals many previously unknown systemic issues, including high variability
	in \ndt test results and systematically under-performing servers in the
	\ookla network.
\end{abstract}

\makeatletter
\let\@authorsaddresses\@empty
\makeatother

\maketitle

\begin{sloppypar}
\section{Introduction}\label{sec:intro} 

Network throughput---colloquially referred to as ``speed''---is among the most
well-established and widely used network performance metrics.  Indeed, ``speed''
is used as the basis for a wide range of purposes, from network troubleshooting
and diagnosis, to policy
advocacy~\cite{ny-iht,battle-for-the-net,penn-rural-mlab,fcc-mba} (e.g., on
issues related to digital equity), to regulation and
litigation~\cite{fcc2022frontier} (e.g., on issues related to ISP advertised
speed).  Given the extent to which stakeholders, from consumers to regulators to
ISPs, all rely on ``speed'', it is in some sense surprising that there is no
consensus on the way to measure it. Absent any standard, many speed tests, varying in
both design and implementation, are used interchangeably. 

Over the past decade, Ookla's Speedtest~\cite{ookla2022speedtest} (``\ookla'')
and Measurement Lab's Network Diagnostic Tool (``NDT'')~\cite{mlab2022speedtest}
have been widely used by both consumers and policymakers: \ookla and NDT report
a daily average of over $10$ million~\cite{ookla2020ntests} and  $6$ million
tests~\cite{mlabs2020bigquery}, respectively. As a result, the compiled datasets
from these two tests, amounting to billions of speed
tests~\cite{ookla2020ntests,clark_measurement_2021}, have become universal
resources for analyzing broadband Internet performance~\cite{battle-for-the-net,
fcc2022frontier, ny-iht, penn-rural-mlab}. Unfortunately, these datasets have
also been used out of context, without a clear understanding of the caveats and
limitations of these tools under different circumstances and
environments~\cite{ookla-ny-case-study}. 

The stakes---and, therefore, the costs---of misuse have also never been higher. In the United
States, Congress has committed \$43.5 billion to Internet infrastructure,
including to last-mile performance and availability
improvements~\cite{infrastructure2022biden}. 
In response,
state and local officials across the country are currently urging consumers to
participate in speed test crowd-sourcing initiatives to help establish which
areas meet the federal funding criteria~\cite{idaho-speed-push}.

To their credit, the organizations who have developed these speed test tools
have tried to prevent misappropriation of the data by issuing guidance about how
the tools and public data should and should not be used. M-Lab has gone as far
as to say that ``\textit{M-Lab's NDT and Ookla's SpeedTest measure fundamentally
different things}''~\cite{mlab-issue-email-1}. While this statement is certainly
true, there has been no study to date about how these differences in tool design
can (and do) yield different results in practice, under different operating
conditions. Acknowledging that \ookla and \ndt are different is, in some sense,
besides the point. 
Although each tool may have been designed with a specific purpose in mind, that
does not mean it can not fulfill---or be appropriated for---other purposes. Such
has been the case with \ndt, which has been used as a tool to measure access ISP
throughput, even though its stated design is to test the throughput of {\em a
single TCP connection}. In light of the significant attention to both of these
tests, it is imperative to develop a rigorous, quantitative, and specific
understanding of the circumstances under which each tool can accurately measure
last-mile speed---and, hence, the context for interpreting each dataset.

To this end, we conduct the first-of-its-kind systematic, comparative study of
\ndt (the latest version of NDT) and \ookla \footnote{We focus on \ookla and \ndt because of   
their popularity with consumers and policy 
makers, but
the method in this paper also applies to other tools.}. We begin 
with a set of in-lab
experiments that allow us to directly compare the tools under controlled network
conditions where ``ground truth'' is known. Next, we conduct more than 80,000
paired wide-area network tests, whereby the two tests are run back-to-back, from
126 home broadband access networks across more than 30 neighborhoods in
one of the largest cities in the United States for nearly a year. In-lab, we use
controlled experiments to characterize how \ndt and \ookla behave under a wide
range of network conditions---specifically, varying throughput, latency, packet
loss, and cross-traffic. We also study how different transport congestion
control algorithms and client types (i.e., browser vs. native client) may affect
the measurements that each tool reports.  Second, we compare the behavior of
these two tools using data from our wide-area network deployment encompassing 10
different ISPs. A unique and important methodological aspect of our study is the
use of \textit{paired speed tests}, where we run \ookla and \ndt in succession.
To our knowledge, this is the first comparative analysis of \ookla and \ndt in
deployment over a significant number of networks for an extended period of time. 

\begin{table}[t]
	\begin{small}
		\begin{tabular}{|p{4.75in}|}
			\hline 
			
			The \ndt client can send at about 95\% of a high-capacity link (up to
			2~Gbps) using only a single TCP connection. This finding updates past work
			that reported a different finding, that a single TCP connection can not achieve a
			throughput approaching full capacity~\cite{feamster2020measuring}. 
			(\cref{subsec:network-conditions}).\\\\
			
			The \ndt client under-reports throughput at higher latencies, in comparison
			to \ookla: The \ookla client reports speeds up to 12\% higher than \ndt at
			200~ms round-trip latency, and up to 56\% higher at 500~ms latency
			(\cref{subsec:network-conditions}).\\\\
			
			Across all households in the wide-area deployment, the median fraction of paired tests 
			for which \ookla
			reports a speed that is 0--5\% higher than \ndt is 73.8\%. The fraction of
			paired tests for which \ookla reports a speed that is 5--25\% higher is
			13.4\% (\cref{subsec:paired-results}).\\\\
			
			For \ookla, the choice of test server can significantly affect the reported
			speed. Tests using certain \ookla servers systematically report speeds 10\%
			lower than other servers. (\cref{subsec:server-selection}). \\\\
			
			\ndt tests are more likely to under-report during peak hours. 43.4\% of
			households observed a statistically significant decrease in \ndt-reported
			download speed tests during peak hours, whereas only 18.9\% of these
			households saw the same for \ookla (\cref{subsec:time-of-day}).\\ \hline
		\end{tabular}
	\end{small}
	\caption{Main results, and where they can be found in the paper.\label{tab:results}}
	\label{tab:main-findings}
\end{table}

Table~\ref{tab:main-findings} summarizes our findings.  We observe significant 
differences in 
Ookla and NDT7’s behavior and explain the causes of these differences. Our results and 
suggestions should help users understand why the reported speed from Ookla and NDT7 may 
differ, as well as guide policymakers towards 
more accurate and appropriate  use of public data sets based on these tools.  To facilitate 
both the  reproduction and extension of  our
results and study, we have released all of the data from this study, as well
as all of the measurement and analysis code we used to conduct the 
study~\cite{netrics-code,netrics-data}. We
view this research as the beginning of a discussion on how to use collective
speed test data to shed more light on the state of broadband Internet access
networks around the United States, and the world.

The rest of the paper is organized as follows. 
Section~\ref{sec:background}
provides background on the design of the \ookla and \ndt speed tests,
including how these tools differ in both design, implementation, and
deployment in practice. Section~\ref{sec:in-lab} describes how we design
in-lab measurements to evaluate the performance of these tools over a
comprehensive set of network conditions. In Section~\ref{sec:netrics}, we
describe a wide-area measurement study of \ookla and \ndt---the first-ever
(and only, to date) comparative study of these two speed test tools in real
deployed networks. Section~\ref{sec:related} describes related work.
Section~\ref{sec:discussion} discusses the implications of our results, including 
guidelines for how our results should (should not) be interpreted and discussion of
possible future directions.

\section{Differences in Speed Test Design}\label{sec:background}

Past research has produced many different approaches to measuring throughput,
available bandwidth, capacity etc. Section~\ref{sec:related} summarizes the
related work in this area, explaining the differences between these approaches.
In recent years, however, \ookla and \ndt have become the two predominant
``speed tests''. Both rely on TCP for transport and attempt to measure the
capacity of the path between the client and server by sending as much traffic as
possible (``saturating'' the link) and computing a throughput value according to
the number of bytes transferred over some sending window. Although both \ookla
and \ndt take this approach, their implementation differs significantly in three
fundamental ways: (1)~the way the client attempts to saturate, or flood, the
link; (2)~the end-to-end path between the client and server; (3)~the sampling
and aggregation of the throughput metrics.

\paragraph{Flooding mechanism.} The mechanisms that \ookla and \ndt use to (or
attempt to) saturate the client-server path are quite different. \ndt opens only
a single TCP connection and runs for a fixed duration, ten seconds. \ookla, on
the other hand, is known to use multiple TCP
connections~\cite{ookla2019throughput}. In our study, we found that \ookla
varies both the number of TCP connections and the test length in response to
changes in the measured throughput over the course of the test (see
Section~\ref{subsec:network-conditions}). The latest versions of both \ookla and
\ndt use TCP websockets. 

\paragraph{End-to-end path.} \ookla and \ndt both measure the throughput of an 
end-to-end network path, which depends on the test server. \ookla and \ndt manage and
operate their server infrastructure in very different ways. Any network can
operate an \ookla server, but servers are typically selected based on client
proximity, as measured by latency. On the other hand, \ndt
servers are owned and operated by a single organization (Measurement Lab). In
addition, \ookla servers are sometimes ``on net'' (within the same ISP as the
client), while \ndt servers are typically ``off-net'' because they reside in
data centers. Connecting to a test server that is off-net means potentially
traversing multiple networks, including transit networks and interconnection
points, that may introduce bottlenecks during a test. In the past, for example,
it has been shown that transit providers such as Cogent have served as
bottlenecks for \ndt tests, and that these transit providers have prioritized
test traffic in times of congestion, thereby affecting the accuracy of the
test~\cite{mlab-issue-email-3}. Section~\ref{subsec:server-selection} discusses
some of these decisions in more detail.

\paragraph{Sampling and aggregation.} The beginning of a TCP connection has a
period called ``slow start'', whereby the client and server transfer data at a
rate that is slower than the steady state transfer rate. Packet loss can also
cause a TCP sender to significantly reduce its sending rate. Such variability,
particularly at the beginning of a transfer, introduces a design choice about
how to sample and aggregate the instantaneous sending rate, as well as where to
sample (i.e. from the sender or receiver). Both \ndt and \ookla report
throughput based on the amount of data transferred in the TCP payload. \ndt
reports the average throughput over the entire test (bytes transferred / test
time). The sampling methodology for \ookla's latest version is not public. The
legacy HTTP-based \ookla client sampled throughput 20 times and discarded the
top 2 and the bottom 25\% samples, thus, effectively disregarding the TCP slow
start period~\cite{ookla-docs}. Although the latest version does not use the
same sampling method (manually verified by inspecting traces), we do observe
some form of sampling whereby it discards low throughput samples.

\section{Controlled Measurements}\label{sec:in-lab}

In this section, we explore how \ookla and \ndt's design and
implementation affect their accuracy in controlled network settings, where we
know and control the ``ground truth'' network conditions. We explore in
particular how these two tools report throughput under a variety of network
conditions, under a range of latency and packet loss conditions, as well as how
the choice of TCP congestion control algorithm affects reported speed. Although we focus on 
\ookla and \ndt, we explore how specific aspects of the test design (e.g., number of 
TCP connections, test duration) 
affect the reported speeds under different network conditions. As such, these results can 
help predict the behavior of other similarly designed speed test tools. 

\subsection{Method and Setup}\label{sec:method} 

For all of our in-lab measurements, we simplify the end-to-end path and
connect the client directly to the server via an ethernet cable. We host both
the \ndt and \ookla server daemon on the same physical server. This setup
allows us to control the network conditions at the client, server, and the
path in between, thus ensuring that any observed differences between
\ookla and \ndt are a result of design and implementation differences between
the tests, as opposed to an artifact of changing network conditions along the
end-to-end path. We use the native speed test clients for both the in-lab
measurements and the wide-area measurements in the next section because the
native client provides metadata (e.g., socket-level information, test UUID)
that the browser version does not provide. In practice, most users run speed tests
through the browser version of these respective tools, despite the fact that
both tools also offer native clients. To understand potential effects of
browser-based measurements, we compare speed test
measurements from both the native and browser clients
(Section~\ref{subsec:usage-modes}); we find no significant difference in the
results reported from each modality.  This finding is a positive and somewhat
surprising result, as early browser-based speed tests did have difficulty
achieving more than 200~Mbps~\cite{feamster2020measuring}.

Finally, because of a bug we discovered in \ndt (discussed in more detail
in Section~\ref{subsec:ndt-upload}), for all \ndt tests we calculate the throughput
of \ndt upload tests using the \texttt{TCPInfo} (based on the throughput
calculated by the server at the kernel level) information, as opposed to the
value reported to the user, which uses \texttt{AppInfo} (based on the sending
rate of the client at the application level).

\paragraph{Hardware.} The server and client are run on identical System76
Meerkat (meer6) desktops (Intel 11th Gen i5 @2.4Ghz, 16-GB DDR
memory, and up to 2.5~Gbps throughput) running Ubuntu 20.04.

\paragraph{Software.} For a measurement server, we used \texttt{ndt-server}
version 0.20.6 and \texttt{\ookla daemon} build 2021-11-30.2159. 
Because there is no publicly available \ookla daemon, we collaborated with the
developers at \ookla to obtain a custom version. For native client
tests, we used \texttt{ndt-client-go} version 0.5.0 and \texttt{\ookla} version
3.7.12.159. For browser-based tests, we use Google Chrome version 100.0.4896.75.
We used the \texttt{tc netem} package to set link capacity, latency, and
packet loss. Both the server and client use the TCP BBR congestion control
algorithm unless otherwise stated.

To facilitate our analysis, we define the \textit{accuracy} as the reported
speed, divided by the true link capacity (i.e., the value that we configure
with {\tt tc}). Despite its name, we do expect this metric to be a fraction
that is always strictly less than 1, due to packet header and protocol
overhead. This metric enables us to compare results across a range of network
conditions, including different link capacities.

\begin{figure}[t] 
	\centering
	\includegraphics[width=0.48\textwidth,keepaspectratio]{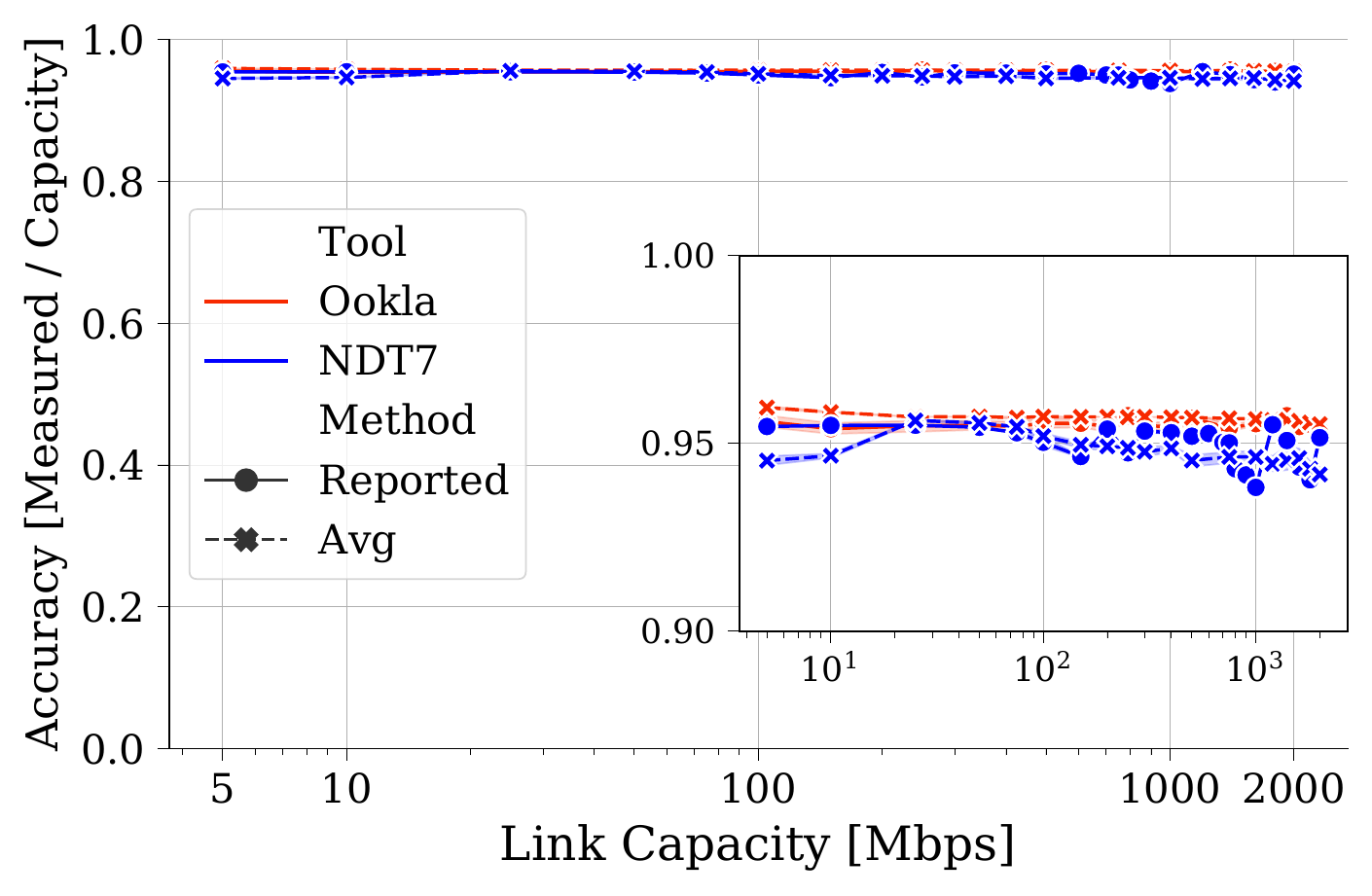}
	\caption{Download accuracy vs. link capacity. Shaded region represents a
	95\% confidence interval for $n=10$ tests. The reported method shows the
	speeds reported by the tool. The average method is the average data transfer
	during the test.} 
	\label{fig:rate-dnld} 
\end{figure}

\subsection{Effects of Network Conditions}\label{subsec:network-conditions} 

We first study how accurately \ndt and \ookla measure the link capacity
under different network conditions. We are interested in quantifying the
effects of both (1)~the mechanisms used to flood the network path and (2)~the sampling
technique used to calculate the final speed. To do so, we show both the
average throughput (calculated using traffic headers) and the final reported
throughput.  The average throughput computation enables us to compare results
from the different approaches, even if the tools report different numbers as a
result of different sampling and aggregation techniques. Having isolated these
effects, we can then separately quantify the effect of sampling technique by
comparing the final reported speed and the average throughput.

\paragraph{Link capacity.} Figure~\ref{fig:rate-dnld} shows each tool's
accuracy when measuring downstream throughput under different link capacities.
The results show that there is no significant difference between \ookla and
\ndt, in terms of both the reported and average data transfer. Both tools can
achieve 95\% link saturation up through a 2~Gbps bandwidth connection using
only a single TCP connection. Past work~\cite{sundaresan2011broadband,
bauer2010understanding} has suggested that a speed test using only a single
TCP connection cannot typically achieve a throughput that approaches link
capacity, typically maxing out at 80\% of the capacity. Our results likely
differ because client operating systems (a bottleneck suggested in previous
work from Bauer~\cite{bauer2010understanding}) have improved and the design of
speed tools has changed. The results in previous studies of
NDT~\cite{sundaresan2011broadband, bauer2010understanding} concern the
original NDT version, which differs from the current implementation of \ndt in
several important ways. Most notably, the original NDT server used TCP Reno,
which often prevented the tests from saturating the
link~\cite{mlab2020evolution}.  The current \ndt server uses TCP
BBR~\cite{mlab2022ndt}.  Absent high latency or packet loss, both tools
achieve similar accuracy across a wide range of link capacities.  Upload
accuracy under different uplink capacities, shown in
Figure~\ref{fig:rate-upld} in the Appendix, show the same trends, with even
less variability.

\begin{figure*}[t]
	\centering
	\begin{subfigure}[t]{0.48\textwidth}
		\centering
		\includegraphics[width=\textwidth,keepaspectratio]{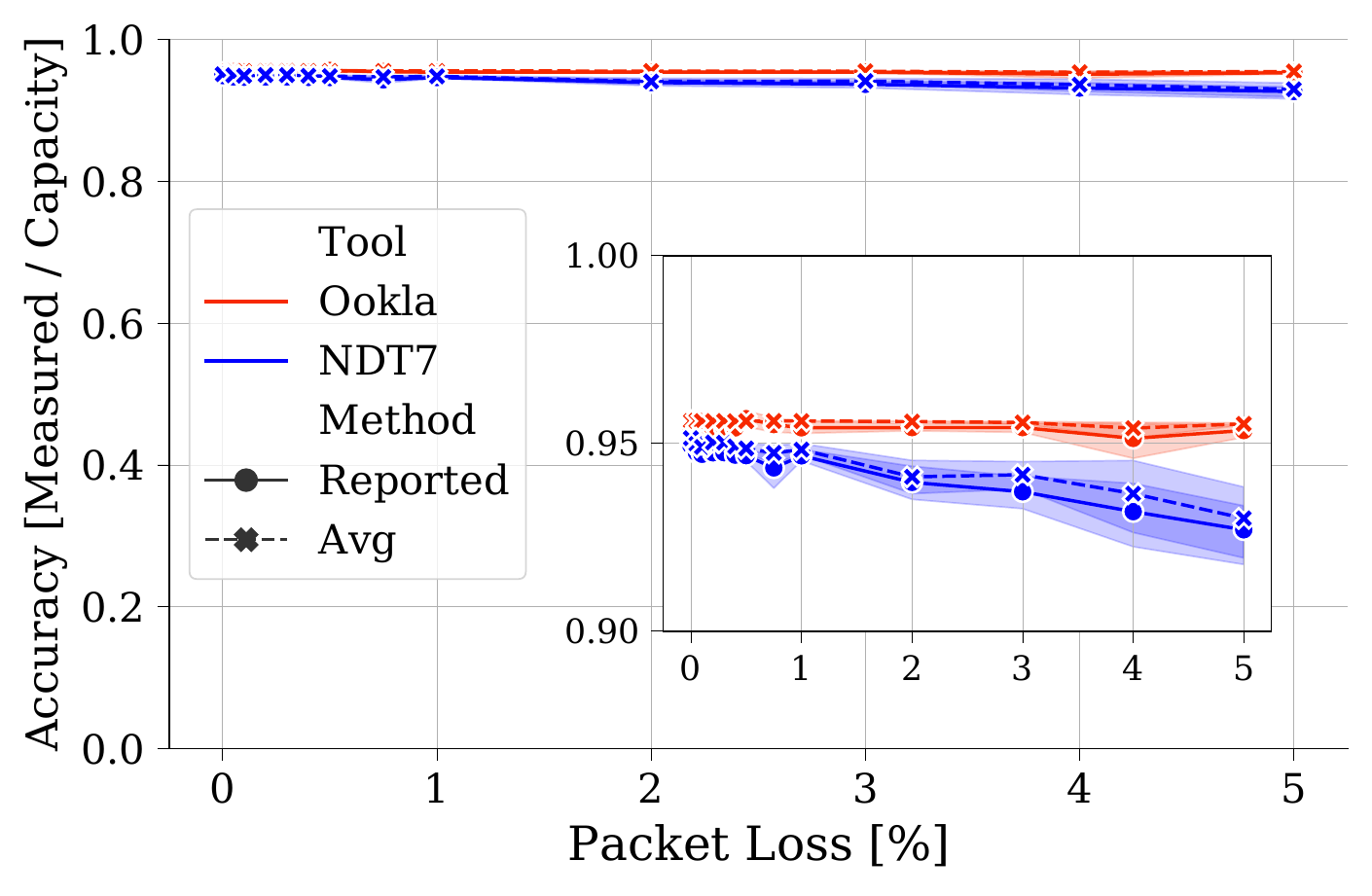}
	\caption{Accuracy vs. Packet Loss}
	\label{subfig:loss-dnld} 	
\end{subfigure}%
	\hfill
	\begin{subfigure}[t]{0.48\textwidth}
		\centering
		\includegraphics[width=\textwidth,keepaspectratio]{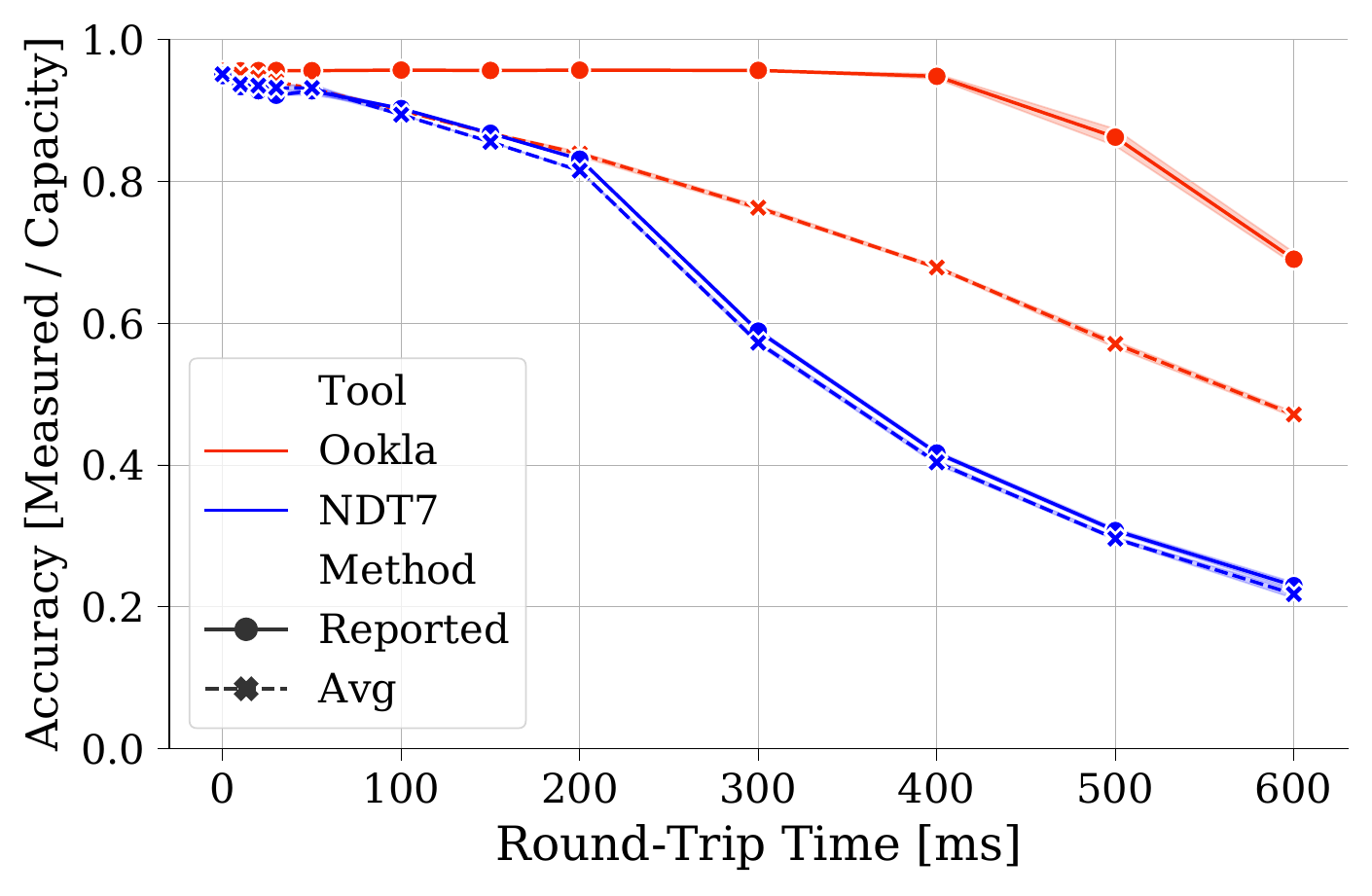}
		\caption{Accuracy vs. Round-Trip Latency} 
		\label{subfig:latency-dnld} 
	\end{subfigure} 
	\caption{Download accuracy under different network conditions. Shaded region
	represents a 95\% confidence interval for $n=10$ tests. The reported method
shows the speeds reported by the tool. The average method is the average data
transfer rate during the test.} 
	\label{fig:latency-loss-dnld}
\end{figure*}

\paragraph{Packet loss.} To study how the packet loss between the server and
client affects the accuracy of \ookla and \ndt, we fix the
uplink and downlink capacity to 100~Mbps and introduce random loss along the path
between the client and server.  For both \ndt and \ookla, packet loss up to 5\% has very little
impact. Figure~\ref{subfig:loss-dnld} shows how download accuracy varies as packet
loss is induced. The range of reported speeds and average throughputs are within
0.3\% for \ookla and 2\% for \ndt. We see little accuracy degradation because
both the server uses TCP BBR as their congestion control algorithm,
which does not use packet loss as a congestion signal.

\begin{table}[t]
	\centering
	\small
	\begin{tabular}{@{}lllrr@{}}
		\toprule
		Rank & Country & Total Tests & > 100ms & > 200ms \\ \midrule
		1    & USA     & 22.9M      & 4.4\%  & 1.5\%   \\
		\rowcolor[HTML]{EFEFEF} 
		2    & India   & 15.7M      & 12.7\% & 4.7\%   \\
		3    & Brazil  & 6.5M       & 22.5\% & 11.5\%  \\ \bottomrule
	\end{tabular}
	\caption{Percentage of \ndt tests conducted in April 2022 with minimum
		round-trip latency exceeding the given threshold.}
	\label{table:ndt-global-latency}
\end{table}

\paragraph{Latency.}  We now study the effects of latency  on speed test
accuracy for a 100 Mbps link capacity.  Figure~\ref{subfig:latency-dnld} shows download 
accuracy is affected as
round-trip time~(RTT) increases. Looking first at reported speed, \ookla is
unaffected by the increase in RTT until round-trip latency exceeds $400$~ms. Meanwhile,
the median speed reported by \ndt decreases to 90\% of the link capacity at
only $100$~ms, further degrades to 83\% at $200$~ms. The difference in
reported speeds is maximum at $500$~ms, 
with \ookla and \ndt reporting a median speed of 87~Mbps and 31~Mbps,
respectively. 

These latency effects are potentially significant because they imply that if a
client test selects a path to a server with high latency, then reported
throughput could be significantly lower---especially in the case of \ndt,
which is more sensitive to high-latency paths. An important question, then, is
how often high-latency tests exist.  To answer this question, we analyze
metadata from recent \ndt tests to see how often \ndt tests are conducted when
the RTT is at least 200~ms. We consider the reported \textit{minimum RTT} for
the tests conducted in April 2022 from the United States, India, and Brazil,
the top three countries in terms of number of tests.
Table~\ref{table:ndt-global-latency} shows the percentage of \ndt tests whose
minimum round-trip latency exceeds 200~ms. All three of these countries see
high latency tests, and in Brazil, 11.5\% of \ndt tests conducted involve
paths where the round-trip latency exceeds 200~ms.  Our in-lab analysis
suggests that these tests would lead to reported speeds of 17\% lower than the
link capacity (or 12\% lower than \ookla) for a 100~Mbps link.

Figure~\ref{subfig:latency-dnld} also illustrates how the choice of a sampling
and aggregation technique to report a throughput result can affect the
robustness of the reported result under high-latency conditions.  As latency
increases, \ookla's average throughput (dashed red line) begins to decrease
while \ookla's reported speed (solid red line) remains constant.  \ookla's
sampling technique discards lower throughput samples, resulting in a reported
throughput that remains high, even as the average throughput achieved by the
test itself decreases.  At higher latency, a TCP connection takes longer to
increase its sending rate---averaging still includes this low-throughput
interval, but \ookla's sampling method discards them and calculates the final
speed using only the higher throughput samples collected towards the end of
the test. 

We investigate why \ookla's average throughput does not degrade similar to
\ndt when RTT exceeds 200~ms. When inspecting the packet traces
from the speed test, we observed that  \ookla adapts both the number of TCP
connections used and the test duration in response to the instantaneous measured
throughput over the course of the test, whereas \ndt does not. \ookla begins to
use multiple TCP connections at $20$~ms of RTT, increasing the number of
connections to as many as eight, at RTT higher than $100$~ms. In contrast,
\ndt only uses a single connection regardless of latency (see
Figure~\ref{fig:latency-nthreads} for details). As for test length, \ookla
begins to increase the median test length from $3.5$ seconds at $0$~ms to median
test length of 11 seconds at $100$~ms---and $15.7$ seconds at $600$~ms, as shown
in Figure~\ref{fig:latency-testlen}. Longer tests allow \ookla's throughput to
get closer to the link capacity than \ndt, whose tests take between $9.5$ and
$10$ seconds, regardless of latency. Past work has suggested that \ookla's test
is not adaptive but instead uses a fixed length~\cite{yang2021fast}; our results
show that the assertion from past work is incorrect, and likely based on
outdated documentation from Ookla~\cite{ookla-old-docs}.

\begin{figure*}[t]
	\centering
	\begin{subfigure}[t]{0.48\textwidth}
		\centering
		\includegraphics[width=\textwidth,keepaspectratio]{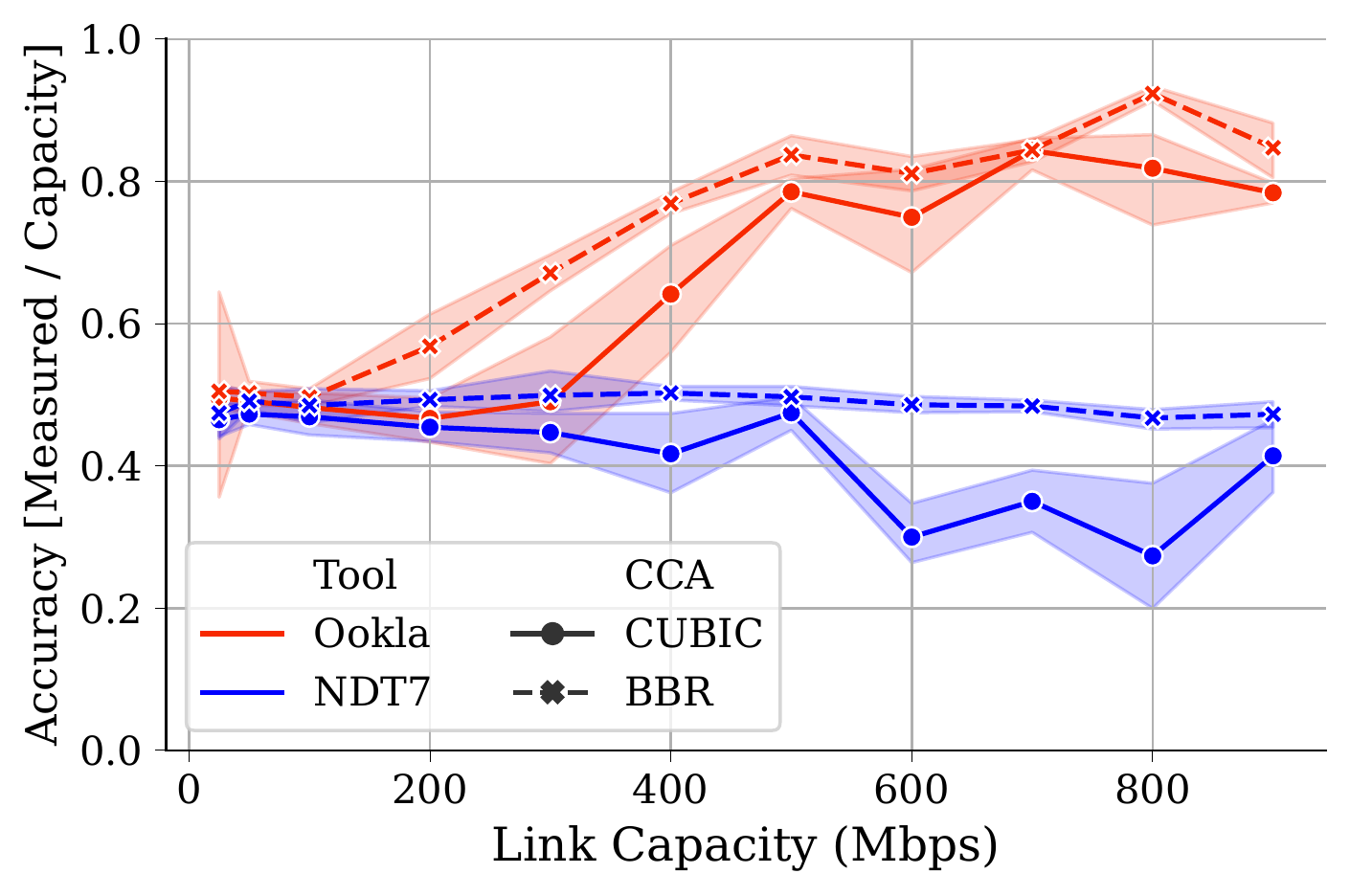}
		\caption{Accuracy vs. Link Capacity}
		\label{subfig:cross-traffic-tput} 	
	\end{subfigure}%
	\hfill
	\begin{subfigure}[t]{0.48\textwidth}
		\centering
		\includegraphics[width=\textwidth,keepaspectratio]{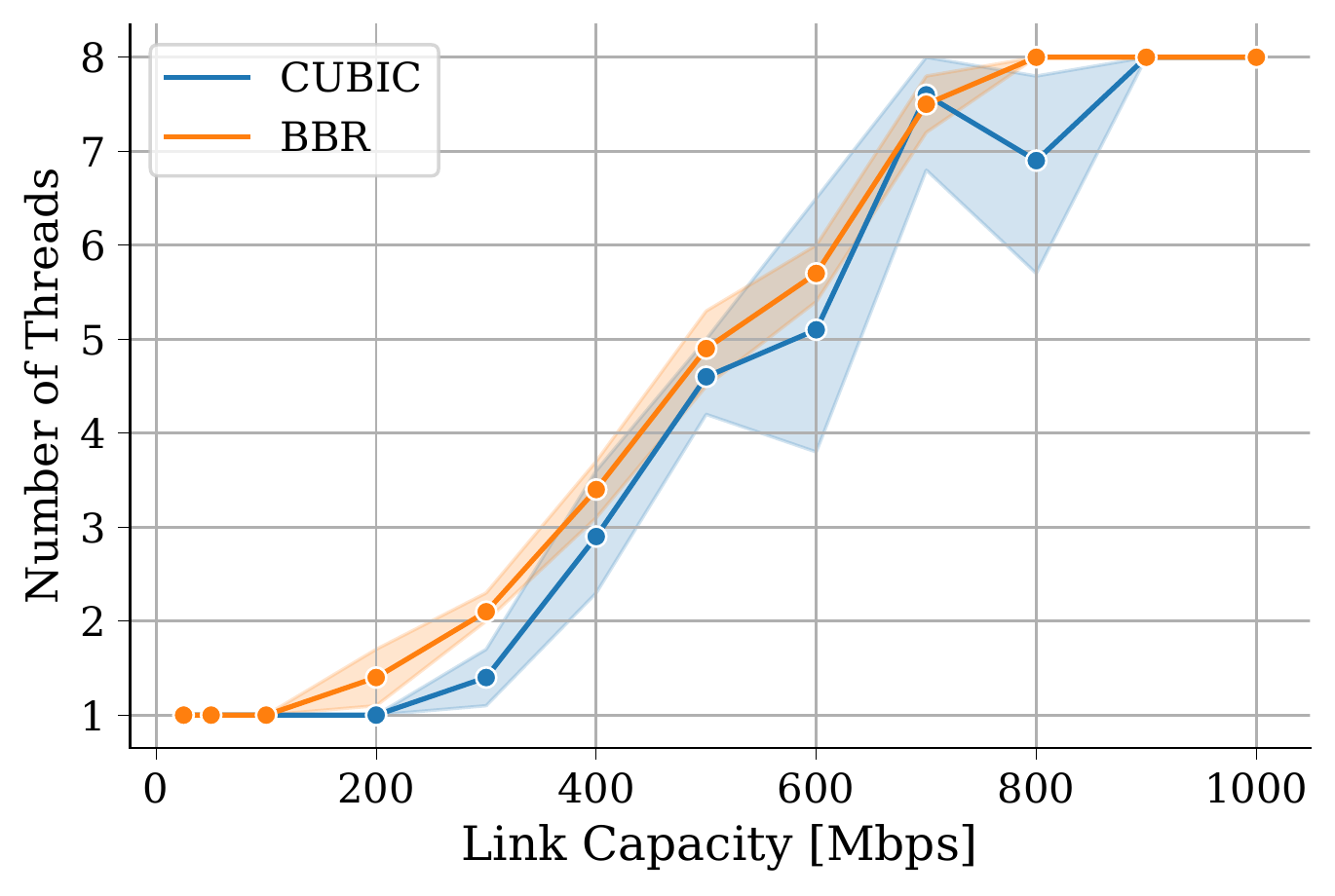}
		\caption{Number of TCP connections opened by \ookla} 
		\label{subfig:nthreads} 
	\end{subfigure} 
	\caption{Download accuracy in the presence of a background TCP Flow. Shaded region
		represents a 95\% confidence interval for $n=10$ tests.} 
	\label{fig:cross-traffic}
\end{figure*}

\vspace{0.5em}
\begin{mdframed}[roundcorner=5pt, backgroundcolor=black!10]
	\textbf{Takeaway}: \ookla and \ndt report similar speeds when the latency
	between the client and server is low. When latency is high (>200ms), \ndt
	and \ookla report speeds that differ by at least 12\% and up to 56\%. \ookla
	is more resilient to high latency because it adapts the number of TCP
	connections and test length, whereas \ndt always uses a single TCP
	connection and has a fixed test length. In addition, \ookla discards low
	throughput samples, leading to an even higher reported speed. 
\end{mdframed}

\paragraph{Cross-Traffic.} Next, we study the effect of background traffic on the
speed reported by each tool. To do so, we run a speed test while a
single-threaded iPerf3 TCP flow runs in the background between the same client
and server machines. We initiate this TCP flow ten seconds before conducting the speed test.
We then repeat this experiment under different link capacities. We set the
buffer size of the bottleneck router to be equal to the bandwidth-delay product
in all of these experiments.

Looking first at \ndt in Figure~\ref{subfig:cross-traffic-tput}, the reported
values are close to half of the link capacity for both TCP BBR and CUBIC. We
might expect this outcome as a consequence of resource sharing--- \ndt uses only a single 
TCP connection and both BBR and CUBIC are
designed to fairly share the link when both flows have the same
RTT~\cite{ha2008cubic,8117540}. We do observe that \ndt reports speeds lower
than $0.5$ when the capacity exceeds 600~Mbps under TCP CUBIC. It is not clear
why \ndt achieves less than its fair share of capacity in these cases. We
suspect it may be due to interactions
between the application and transport layer.  In contrast, \ookla using 
TCP CUBIC reports a fair share ($0.5$) up to 300~Mbps, but reports values as high
as 80\% of the capacity when link capacity is greater than 500~Mbps. We see a
similar trend when \ookla uses TCP BBR, the only difference being that the reported speeds
become greater than the fair share at lower speeds, i.e., at 200~Mbps. Upon further
inspection, we find that \ookla starts using more than one TCP connection when
the underlying link capacity is high (see Figure~\ref{subfig:nthreads}). This
behavior appears to depend on the link capacity and differs slightly for
the two TCP congestion control algorithms. \ookla using TCP BBR starts opening
multiple connections at a lower link capacity than it does when using TCP
CUBIC, which explains why
the reported values are higher for TCP BBR. We did not find any trends
corresponding to test
duration, indicating the higher speed values are mostly due to the use of
multiple TCP connections, not length of test.

Our finding that \ookla and \ndt behave very differently in the presence of
cross-traffic raises an interesting and important discussion about the
complexity of speed measurements and the ``correct'' way to characterize the
results, especially given that many crowd-sourced measurements likely
experience some amount of cross-traffic. Specifically, \ndt reports a single
connection's fair share of capacity, whereas \ookla reports the results of a
more aggressive approach that relies on multiple parallel connections.  It is
thus critical to interpret the measurements from these tools in context,
because some applications (e.g., web browsing) tend to behave more like \ookla,
whereas others (e.g., simple file transfer) may behave more like \ndt.  In this
vein, greater transparency about how these tools are designed---and what the
measurement results likely reflect, in context, as we have done in this
paper---would almost certainly be beneficial to anyone who uses these tools or
the resulting data.

\vspace{1em}
\begin{mdframed}[roundcorner=5pt, backgroundcolor=black!10]
	\textbf{Takeaway}: Under cross-traffic, \ndt reports its fair-share throughput while 
	\ookla is more aggressive, reporting values up to 90\% of the link capacity. The 
	difference can be attributed to the use of multiple TCP connections by \ookla. While 
	the question of\textit{ which tool is more``accurate''} is complex, bringing 
	transparency into the test methodology such as the number of TCP connections used 
	can better help users to interpret the test results. 
\end{mdframed}

\begin{figure}[t]
	\centering
	\begin{subfigure}[]{0.48\textwidth}
		\centering
		\includegraphics[width=\textwidth,keepaspectratio]{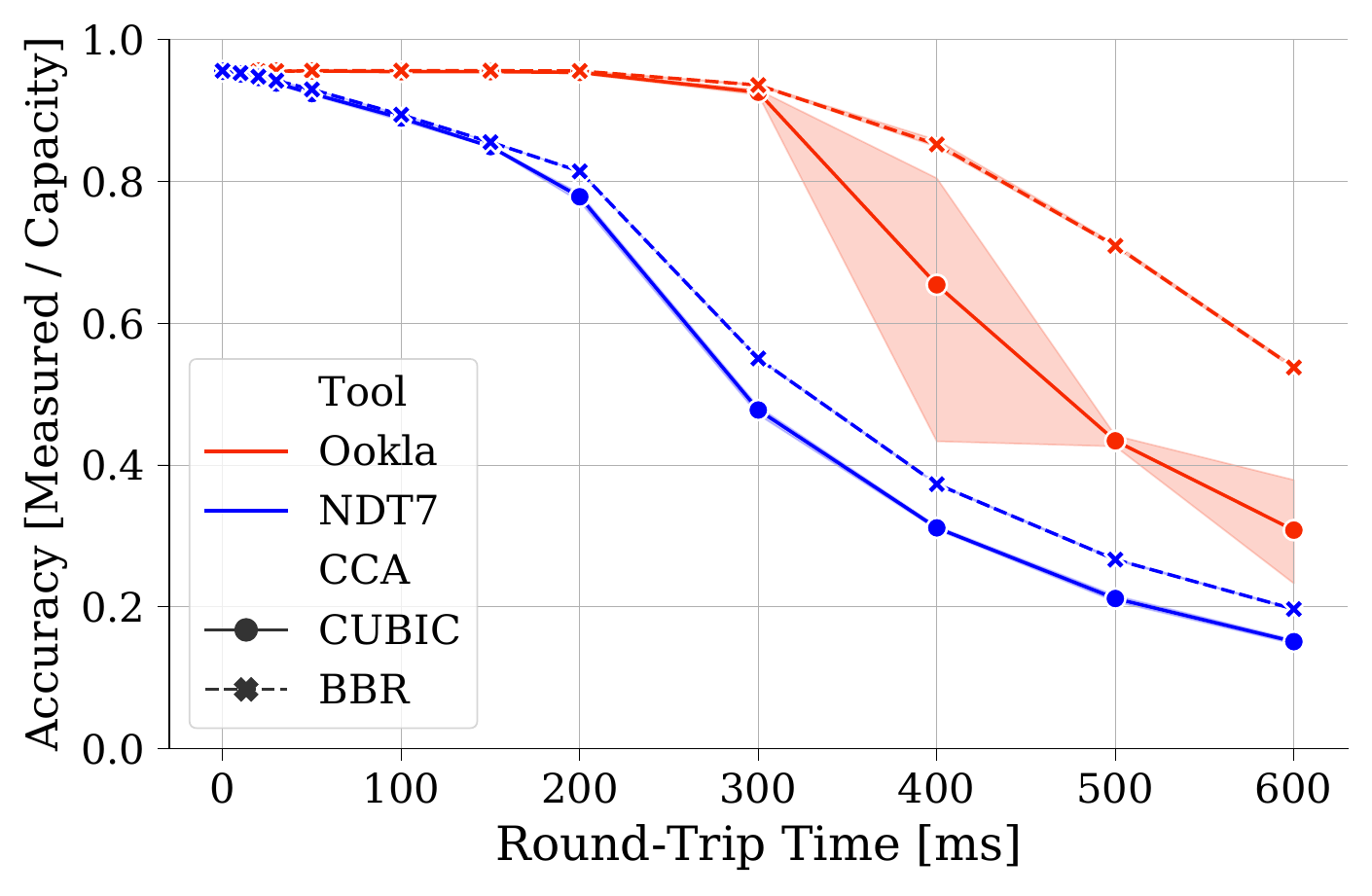}
		\caption{Accuracy vs. Latency}
		\label{subfig:cca-up-latency}
	\end{subfigure}%
	\hfill
	\begin{subfigure}[]{0.48\textwidth}
		\centering
		\includegraphics[width=\textwidth]{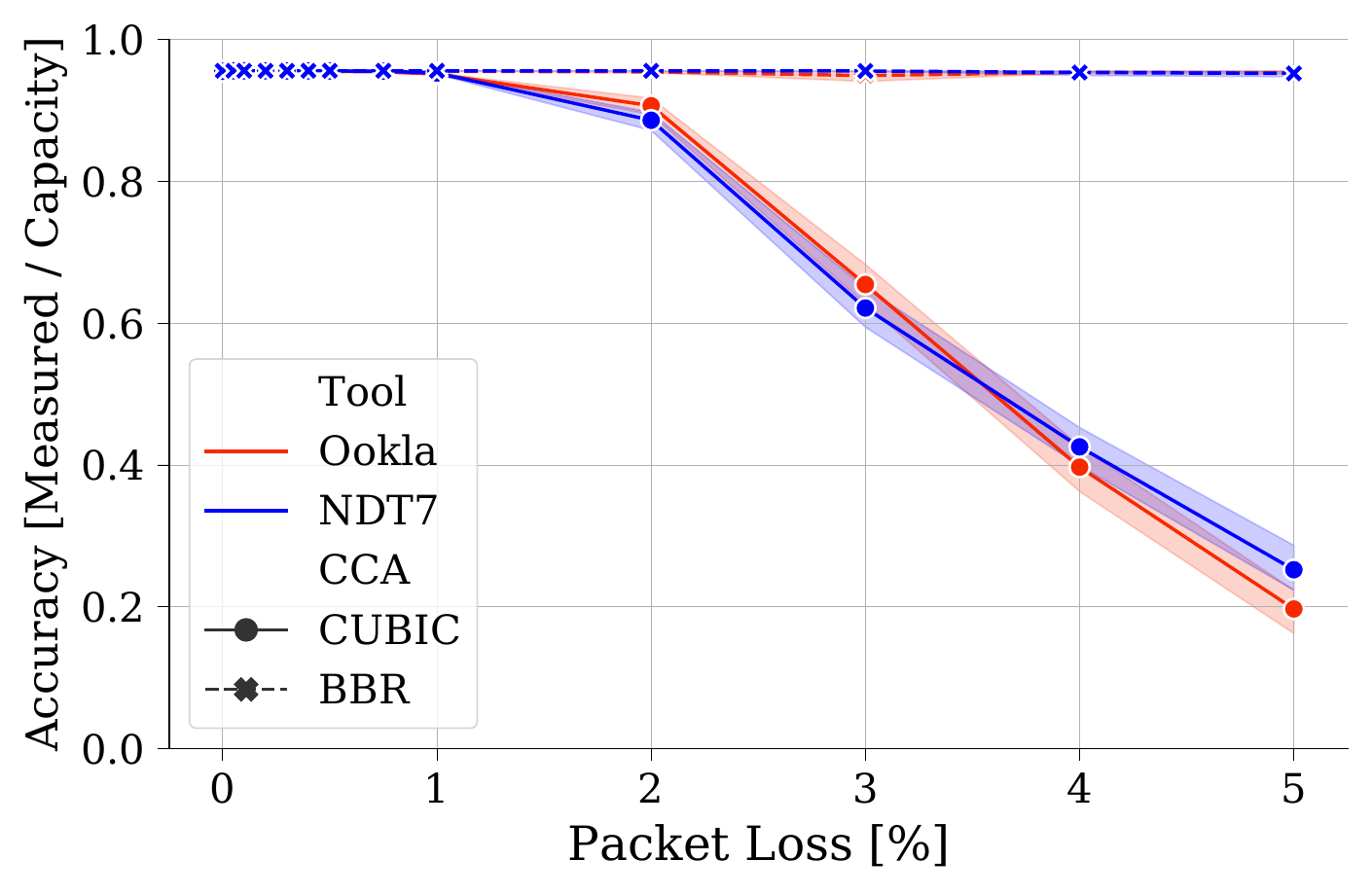}
		\caption{Accuracy vs. Packet Loss}
		\label{subfig:cca-up-loss}
	\end{subfigure}%
	\hfill
	\caption{Tool accuracy under different client-side TCP congestion control
		algorithms for upload tests. Shaded regions represent a 95\% confidence
		interval for n=10 tests.}
	\label{fig:cca-up}
\end{figure}

\begin{figure*}[h]
	\begin{subfigure}[t]{0.33\textwidth}
		\centering
		\includegraphics[width=\textwidth,keepaspectratio]{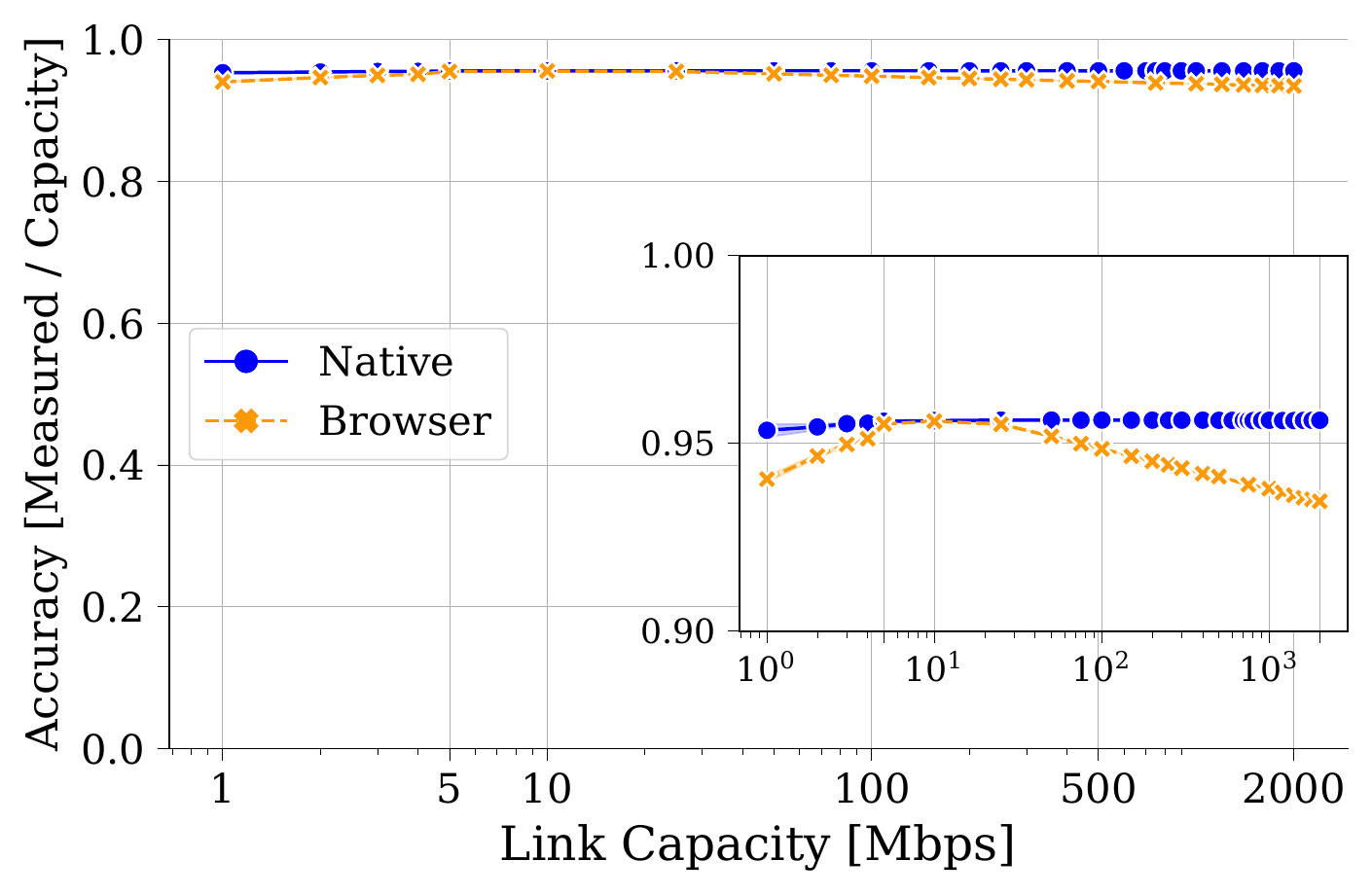}
		\caption{Link Capacity}
		\label{subfig:browser-rate-up}
	\end{subfigure}%
	\hfill
	\begin{subfigure}[t]{0.33\textwidth}
		\centering
		\includegraphics[width=\textwidth]{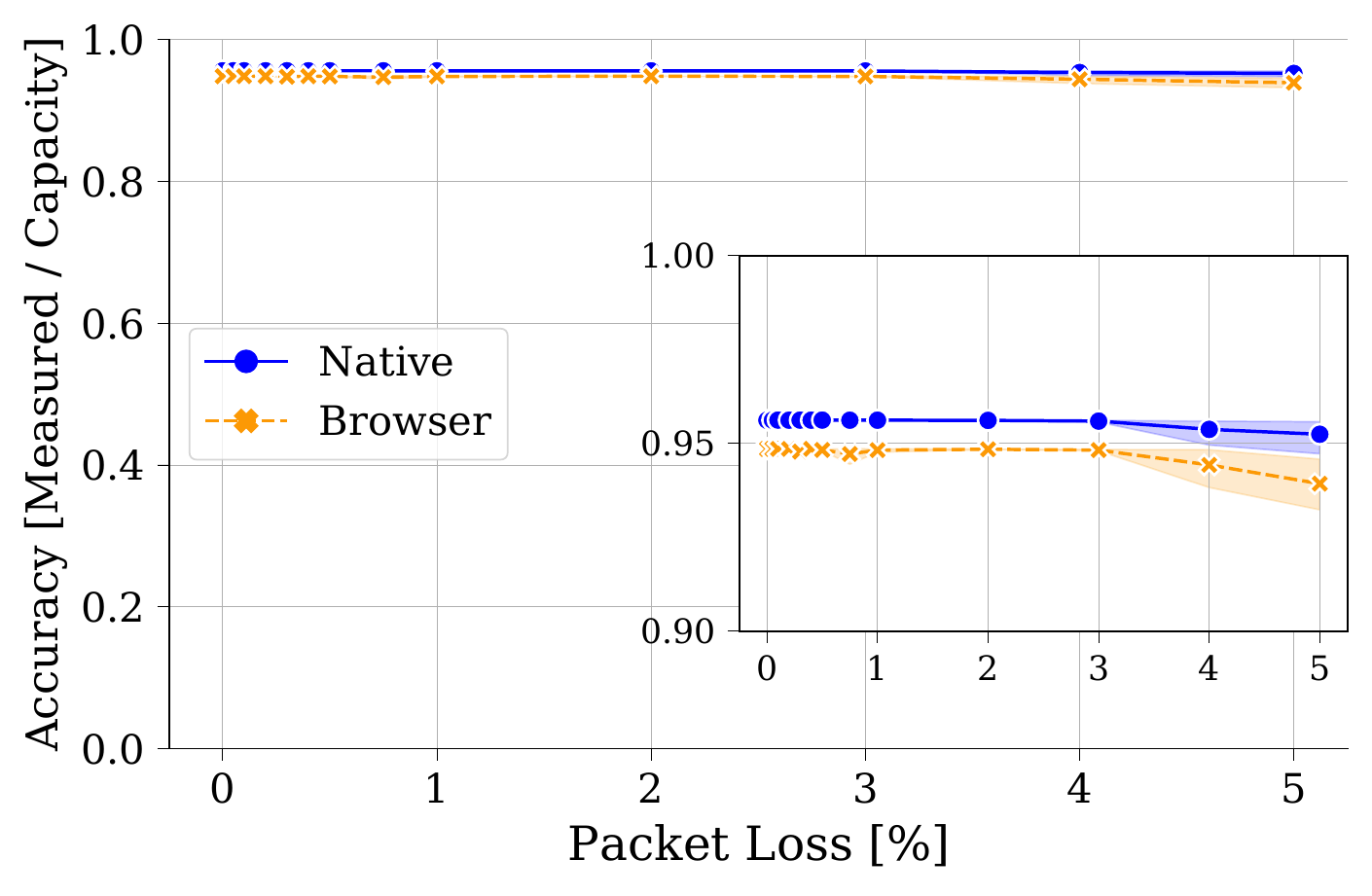}
		\caption{Packet Loss}
		\label{subfig:browser-loss-up}
	\end{subfigure}%
	\hfill
	\begin{subfigure}[t]{0.33\textwidth}
		\centering
		\includegraphics[width=\textwidth]{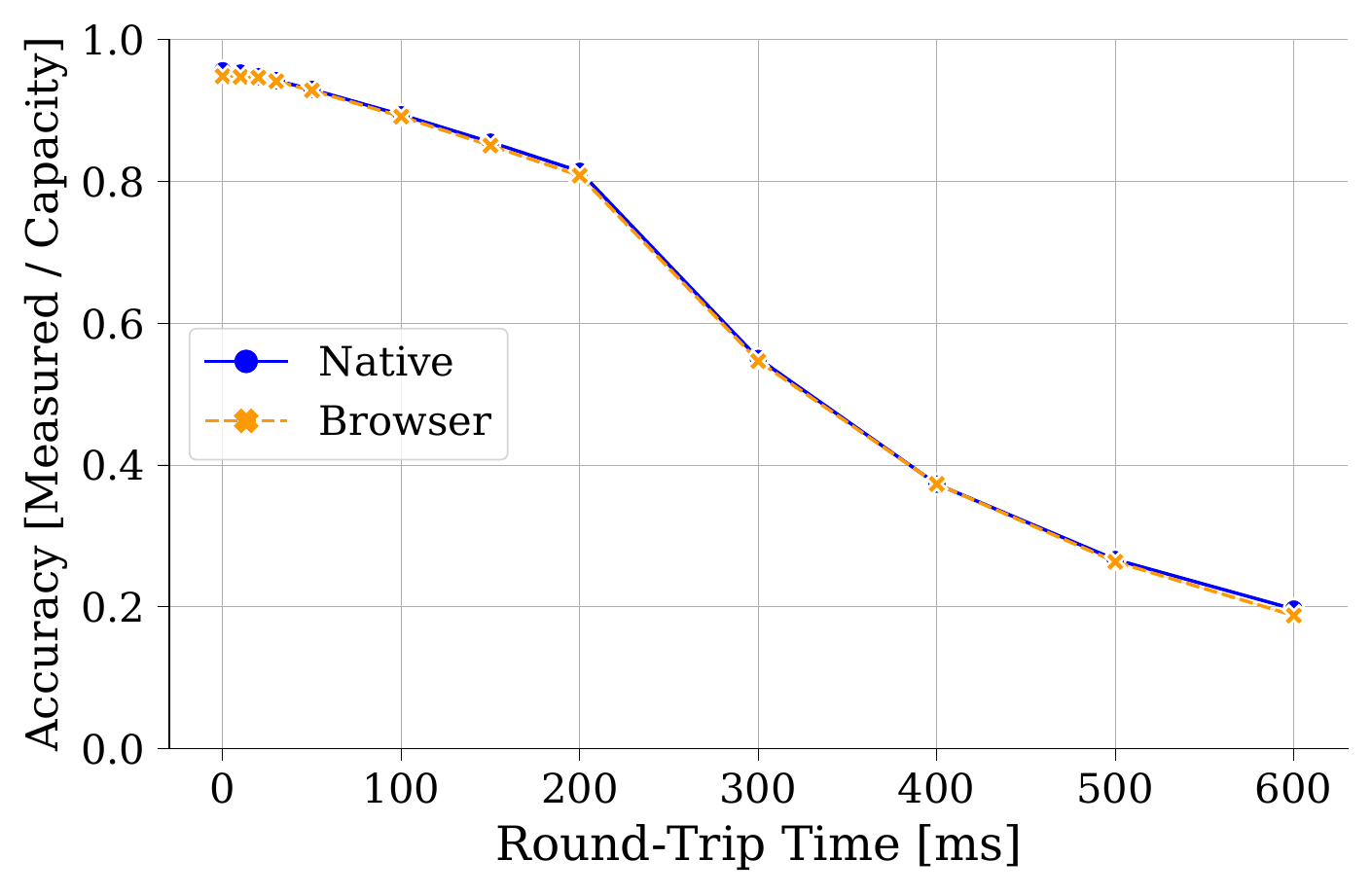}
		\caption{Latency}
		\label{subfig:browser-latency-up}
	\end{subfigure}
	\caption{\ndt accuracy on different client types (browser vs. native) for
		upload tests.}
	\label{fig:browser-up}
\end{figure*}

\subsection{Effects of Client Modalities}\label{subsec:usage-modes}

In addition to variable network conditions, client-side usage
modalities can affect the accuracy of \ookla and \ndt. We study the
effect of two such modes: (1)~choice of TCP congestion control algorithm (CCA)
and (2)~choice of client types (browser vs. native). 

\subsubsection{Congestion Control Algorithm}
We compare two commonly used TCP CCAs--- CCAs that have been used in
these speed tests---TCP CUBIC and TCP BBR. We choose these
algorithms because TCP CUBIC is the default CCA on all current Linux, Windows,
and Mac machines while TCP BBR is the CCA recommended for the \ndt test. In
addition, TCP CUBIC and TCP BBR use different congestion signals, packet loss
and changes in latency, respectively. In this part of our study, we 
study only upsteam throughput measurements, as the choice of client-side CCA only affects the
sending rate. 

\paragraph{Latency.} Figure~\ref{subfig:cca-up-latency} shows the reported
speed at the indicated round-trip times for different client-side CCAs. The
difference in average speed between clients running TCP BBR and TCP CUBIC can be up to
27\% of the link capacity for \ookla, and as much as 7\% of the link capacity
for \ndt. The greater discrepancy for \ookla is caused by \ookla opening
multiple TCP connections when using TCP BBR but not for TCP CUBIC. It is unclear why
the choice of CCA causes this behavior. 

\paragraph{Loss.} When using TCP CUBIC, both \ookla and \ndt experience
severely reduced
accuracy at packet loss rates at 2\% and higher.
Figure~\ref{subfig:cca-up-loss} shows how \ookla and \ndt accurately measure
link capacity under different CCAs and packet loss regimes. As expected, clients
running TCP CUBIC suffer because CUBIC uses packet loss as a congestion signal
and slows it's sending rate, whereas TCP BBR is resilient to losses. 

\subsubsection{Client Type}

In this section, we compare the accuracy of the \ndt browser client and the
\ndt native (command-line) client. We perform these 
experiments for two reasons: (1)~
in the past,
browser-based tests have been less accurate at high link
capacities~\cite{feamster2020measuring}; (2)~it is likely
that most speed tests are conducted using the browser. 
There is no readily available way to
conduct the same comparison for \ookla because it is not open-source, and 
we cannot configure the test server in \ookla's current browser client.

Figure~\ref{fig:browser-up} shows the \ndt upload accuracy for both the
native and browser client as the network conditions vary. Recall
that we calculate speed using the \textit{TCPInfo} messages for both the native
and browser client. Overall, there is very little difference in accuracy 
between the two client types. As packet loss and latency between the client and
server is induced, the difference in median accuracy between the browser and
the native client is within 1\%. Although there is a small dip in accuracy as
the link capacity is increased, the difference is less than 2\%. 

\vspace{0.5em}
\begin{mdframed}[roundcorner=5pt, backgroundcolor=black!10]
	\textbf{Takeaway}: When using TCP BBR, both \ndt and \ookla are more
	resilient to increases in latency and packet loss than when using TCP CUBIC.
	There is no significant difference between the \ndt browser client and the
	\ndt native client.
\end{mdframed}

\begin{figure}[t] 
	\centering
	\includegraphics[width=0.5\textwidth,keepaspectratio]{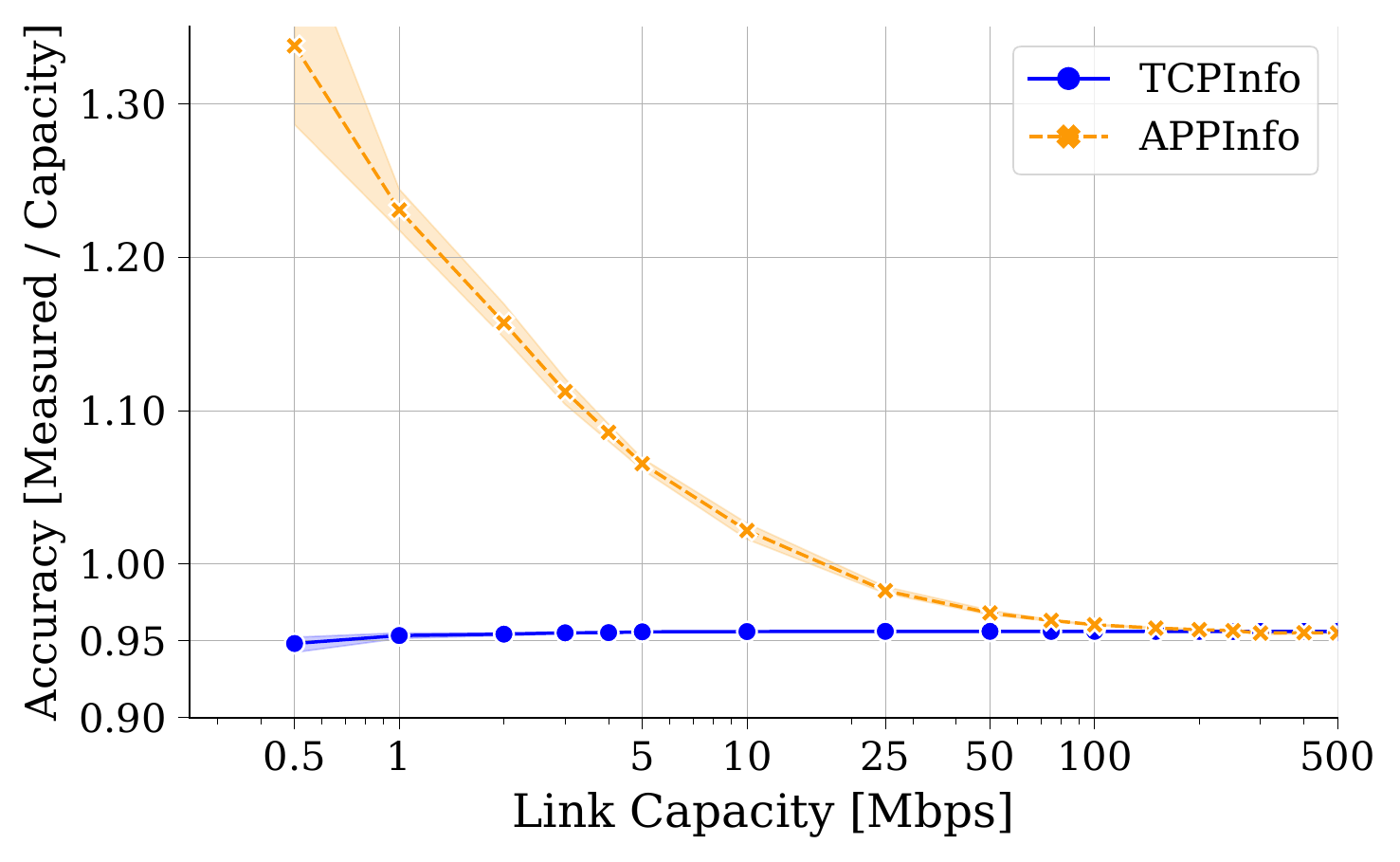}
	\caption{\ndt upload throughput calculated using \texttt{TCPInfo} messages
	and \texttt{AppInfo} messages. Note that the y-axis begins at 0.9. Shaded
	region represents a 95\% confidence interval for $n=10$ tests.}  
	\label{fig:ndt-upload-method} 
\end{figure}

\subsection{\ndt Upload Test Inconsistencies}\label{subsec:ndt-upload}
Our controlled experiments revealed that the \ndt native client would report
upload speeds exceeding the link capacity by up to 34\%. Upon making this
observation, we studied how the \ndt native client calculates the final reported
speed. The remaining  section discusses how we isolated the inconsistency,
identified the bug in \ndt, and communicated the finding (and fix) to the
Measurement Lab team.

During an \ndt test, the client and server exchange two types of messages:
\texttt{AppInfo} messages and \texttt{TCPInfo} messages. The \texttt{AppInfo}
messages track the amount of data transferred at the application level, while
the \texttt{TCPInfo} messages track the amount of data transferred at the
kernel-level. Before fixing the bug, the native \texttt{ndt-go-client} used the
\texttt{AppInfo} messages to calculate speed. This choice only affects upload
results, as this method counts all bytes sent to the TCP socket, as opposed to
bytes that are successfully sent across the network. Because download speed is
also calculated at the client, only successfully received bytes are counted, so
the calculation method does not affect the reported download speed.

Figure~\ref{fig:ndt-upload-method} shows how \ndt upload accuracy varies
across different link capacities when calculated using the \texttt{AppInfo}
and \texttt{TCPInfo} messages. The native \ndt client reports a
throughput greater than the link capacity at link capacities under 10~Mbps. At 
0.5~Mbps, the native \ndt client reports a speed that is 134\%
of the link capacity. It would thus be inappropriate to {\tt AppInfo} to
measure upstream throughput. This discrepancy is of special concern for
measuring upstream
throughput on residential networks, where it is common for the provisioned uplink
capacity to be under 10~Mbps. In light of this finding, we also verified how the
\ndt browser calculated the final speed. To our surprise, we found 
that the browser calculates
upload speed using the \texttt{TCPInfo} messages, not the \texttt{AppInfo}
messages.  

We communicated our findings to the \ndt team at Measurement Lab, who have
already implemented a bug fix for the \ndt native client to use the
\texttt{TCPInfo} messages instead of the \texttt{AppInfo} messages to compute
the speed. Because the \ndt native client sends data in units of size up to
1~MB, the number of transferred bytes at the sender and receiver can differ by
up to 1~MB. This characteristic explains why tests at lower link capacities
exhibited greater relative differences than tests at higher link capacities.
Although this bug is fixed now, it has implications for
analysis of past data. Since many \ndt native client tests were conducted when speed was
calculated using the \texttt{AppInfo} messages, \textit{we suggest to recalculate the upload 
speeds using the \texttt{TCPInfo}
statistics available in the Measurement Labs BigQuery
database when analyzing past data~\cite{mlabs2020bigquery}}.  
\vspace{0.5em}
\begin{mdframed}[roundcorner=5pt, backgroundcolor=black!10]
	\textbf{Takeaway}: \ndt upload tests conducted using the native client
	report speeds greater than the link capacity for link capacities under
	10~Mbps. This overestimation is caused by the native client reporting
	sender-side statistics. This inaccuracy is especially important to
    consider in the context of
    residential networks, for which upload speeds (especially for historical
    data) may be less than 10~Mbps.
\end{mdframed}

\section{Wide-Area Measurements}\label{sec:netrics}

We complement our in-lab analysis with a longitudinal study of \ookla and \ndt
on a set of diverse, real-world residential networks. In this study, we conduct
at least daily \ookla and \ndt speed tests over a 9-month period. We begin by
asking ``\textit{how often do \ookla and \ndt report different speeds and to
	what extent?}''. To answer this question, we compare the reported speeds
	from the paired \ookla and \ndt tests. Although differences in test design (e.g. sampling 
	mechanisms) might explain
	small differences in reported speed , they do not explain larger differences. To this end, we
	investigate how differences in the \ookla and \ndt client-server network
	path might cause these larger differences. 

\subsection{Method and Setup} 

We deploy Raspberry Pis (RPis) in $126$ households, spanning $10$ ISPs from
November 2021 to August 2022.   All the households are located in a major city. A focused 
geography enables us to measure 
the same subset of  \ookla or
\ndt server infrastructure from multiple vantage points; thus, enabling
discovery of server-side issues in Section~\ref{subsec:server-selection}. We recruited these 
households with the help
from various community organizations, as well as our own local outreach
initiatives. We made significant efforts to minimize bias by ensuring our sample 
size spans households from different neighborhoods, ISPs, and speed tiers.

We use Raspberry Pi
4 Model B devices (Quad-core Cortex-A72 CPU, 8GB SDRAM, and up to 1~Gbps
throughput). Each RPi conducts a series of active measurements, including at
least daily \ookla and \ndt speed tests at random times of day. We eliminate any
WiFi effects and conduct the speed test directly by connecting the RPi to the
network router via Ethernet. We verify that the speed test results are not
limited by the RPi hardware by performing in-lab tests comparing the RPis
against the desktop machines used for controlled experiments. We see no significant 
performance 
difference between RPis and desktop machines
 for download tests. For upload tests, we do observe that, for link
capacities of 900~Mbps and 1~Gbps, \ndt tests conducted on the RPi are on
average 5\% lower than those conducted on the desktop. \ookla reports $5\%$ lower
speeds at 1~Gbps on the RPi than on the desktop. Figure~\ref{fig:hardware-rate}
shows this comparison at various link capacities. These discrepancies should not
affect our conclusions at the speeds we test because (1)~most ISP offerings for
residential networks are significantly lower than 1~Gbps (e.g., up to 35~Mbps
for Comcast DOCSIS Cable); (2)~our analysis compares \ookla and \ndt speeds
relative to each other, and both tools degrade similarly due to RPi except at
speeds above 900~Mbps. 

\paragraph{Pairing \ookla and \ndt tests.} To isolate the effects of \ookla and
\ndt's server infrastructures, we must hold all other test conditions constant,
meaning we must control for the network conditions present along the parts of
the end-to-end path that \ookla and \ndt tests have in common. Because network
load can vary over time, comparing \ookla and \ndt test results taken from
different points in time would be to inaccurately compare the network in two
different states. As such, we conduct \textit{paired speed tests}, whereby we
run \ookla and \ndt tests back to back. This approach ensures that the network
conditions along the shared portion of the end-to-end path are reasonably
similar during both tests. Running the tests back to back may introduce bias as the first test 
may starve any background traffic giving advantages to the second test . However, we find 
that is not the case through in-lab experiments. This is because the first test runs both 
download and upload test followed by a period of minimal network activity when the second 
test is picking a server. Thus, there is enough time gap between download tests of the two 
tools for any 
persistent background flow to 
recover.

\paragraph{Normalizing speed.}  We surveyed participants about their speed tiers 
in the 
beginning 
of the study. However, we found discrepancies in the reported speed tier and the 
measured 
speeds in a few cases. Moreover, a few
participants changed their Internet service plan during the course of the study, resulting in 
different download and/or 
upload 
speeds. Given the lack of ground truth speed tier, we rely on measured speeds. 
Specifically, we define the \textit{nominal speed} 
for a
given household to be the 95th percentile speed across all speed tests of a
given tool from that household. We choose the 95th percentile in accordance with
past work that studied Internet performance in residential
networks~\cite{sundaresan2011broadband}. Using the nominal speed, we compute the
normalized speed for each test $i$ as follows:

\begin{equation}\label{eq:normalized-thruput}
	\hat{S}_{i} = \frac{S_{i}}{S_{95th}}
\end{equation}
\noindent
where $S_{i}$ is the speed reported by test $i$ and $S_{95th}$ is the
$95$th percentile result across all speed tests from that particular
household. In addition to using the nominal speed to normalize test results, we
use it when assigning households to different speed tiers. 

\paragraph{Tier changes.} We identify speed tier changes by
manually inspecting each household's speed test results over time. To
accommodate speed tier changes in our study, we treat measurements taken before and
after the speed tier upgrade as two distinct households. In total, 
we observe $23$ instances of download speed tier change and $20$ instances of upload speed 
tier 
change. Therefore, we end up with a different number of households for download and
upload tests: $135$ and $132$, respectively. 

\paragraph{Characterizing test frequency.} For download tests, there is a
median of 336 paired tests across households, with a minimum of 32 and maximum
of 3,381. For upload tests, there is a median of 347 paired tests, minimum of
32 and maximum of 3,381 for each household. The CDF of the number of tests is
shown in Figure~\ref{fig:ntests}. For each household, we run \ookla and \ndt
tests at least once daily for an average of 128 days and median of 108 days. As
many households we study have ISP-imposed data caps, we set a limit on the
monthly data consumption our tests can use. Households with higher speed
connections will consume more data per test and thus will run fewer tests than
household with lower speed connections. Over the course of our deployment,
high-speed (> 500 Mbps download) households have a median of 241 paired tests
while low-speed (< 500 Mbps download) households have a median of 640. 

\subsection{Comparing Paired Speed Tests}\label{subsec:paired-results}
Conducting paired tests allows us to directly compare \ookla and \ndt results
because both tests were run under similar access link conditions. In this
section, we characterize the differences in reported speed for each paired test.
We consider the following questions: (1)~for each household, is
the average difference in reported speeds over time statistically significant
and (2)~for each paired test, how often and to what extent do the reported speeds
differ?

\subsubsection{How often do the average test results differ?}
\label{subsec:t-test-paired}\hfill\\ 

\paragraph{Method.} Using longitudinal measurements from each household, we analyze
whether the average difference between the speed reported by \ookla and \ndt is
statistically significant. We leverage the fact that we have paired observations
and test for significance using a paired t-test. Our sample satisfies
the assumptions required to conduct paired t-test as follows: (1)~the measurements
are collected at random time intervals, (2)~the \ookla and \ndt tests are
conducted back-to-back, creating a natural pairing, and (3)~the number of paired
observations per network is great enough (> 30) to satisfy the normality
condition~\cite{peck_olsen_devore_2009}.  

The null hypothesis, $H_{0}$, is that for a given household, the average
difference between each pair of \ookla and \ndt tests is 0. For each household, we
then compute the p-value, or the probability of observed values under the null
hypothesis. Using a significance level $\alpha=0.01$, we then reject $H_{0}$ if
the calculated p-value is less than $\alpha$.  We choose $\alpha=0.01$ based on
the norms in other fields where hypothesis testing is more common. Choosing
$\alpha=0.01$ can be interpreted as there being a 1\% chance of rejecting
$H_{0}$, when $H_{0}$ is true~\cite{peck_olsen_devore_2009}. 

\begin{figure}[t]
		\centering
		\includegraphics[width=0.75\textwidth,keepaspectratio]{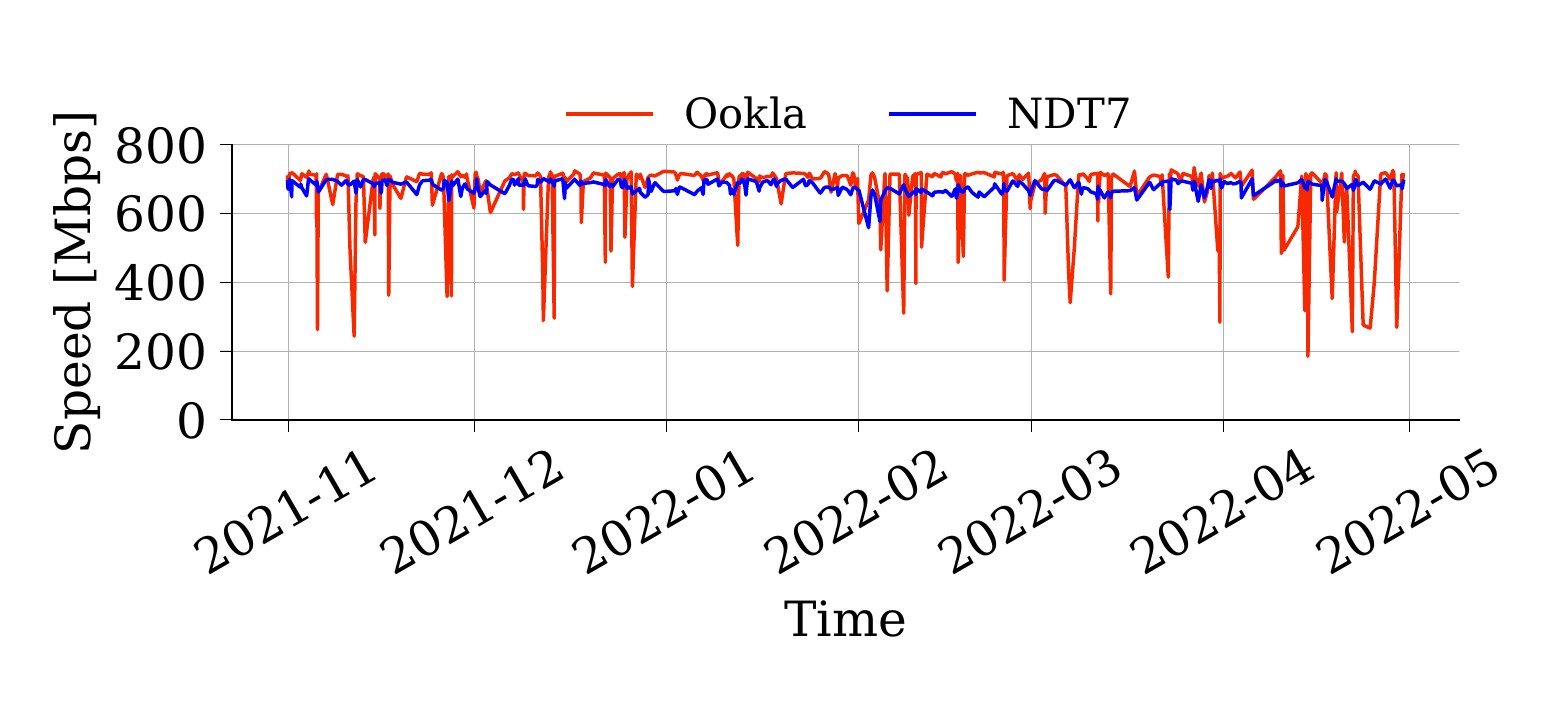}
		\captionsetup{width=\textwidth}
		\caption{Download speed test results over time from a single household:
		\ookla results show high variability with occasional dips in reported
		speed.}
		\label{fig:similar}
\end{figure}

\paragraph{Results.} For download tests, we can reject $H_{0}$ for 82.2\% of
households (111 out of 135) and can conclude that the average difference in
reported speeds is statistically significant for those 111 households. Turning
next to upload tests, we can reject the null hypothesis for only 59.8\% (79 out
of 132) households. This is likely because 72\% of households have much lower
upload speeds (< 50~Mbps). For these lower upload speed households, the ISP
access link is likely the bottleneck. As observed in our in-lab tests, both
tools report similar values under low bottleneck link capacity.

Focusing now exclusively on households for which we could reject $H_{0}$, we
calculate the mean difference between \ookla and \ndt speeds for these
households. For download tests, the mean difference is within 5\% (10\%) of the maximum 
test mean for 75 (91) out of 111 households. Similarly, for upload tests, the mean difference 
is within 5\% (10\%) of the maximum
test mean for 50 (61)
out of 79 households. Thus, overall, the average differences between \ookla and \ndt are 
either not 
statistically significant or within 5\% (when they are statistically 
significant) for 73\% and 78\% households, for download and upload tests, respectively.

In comparison to average differences, we find that the difference within individual paired test 
can be
high. Figure~\ref{fig:similar} illustrates the download speeds reported by \ookla and \ndt tests
over time for a single household, for which the average difference between \ndt
and \ookla results was not statistically significant (p-value = $0.021$). For
$80\%$ of paired tests from this household, \ndt reports a lower speed (up to
80~Mbps lower) than \ookla. However, \ookla occasionally reports a much lower
speed (up to 400~Mbps lower), bringing the \ookla average speed closer to that
of \ndt. Given this range in differences within a single paired test, we next
study the magnitude and frequency of these differences within each paired test.

\begin{figure*}[t]
	\centering
	\begin{subfigure}[t]{.48\textwidth} \centering
	\captionsetup{width=.5\linewidth}
	\includegraphics[width=1\textwidth,keepaspectratio]{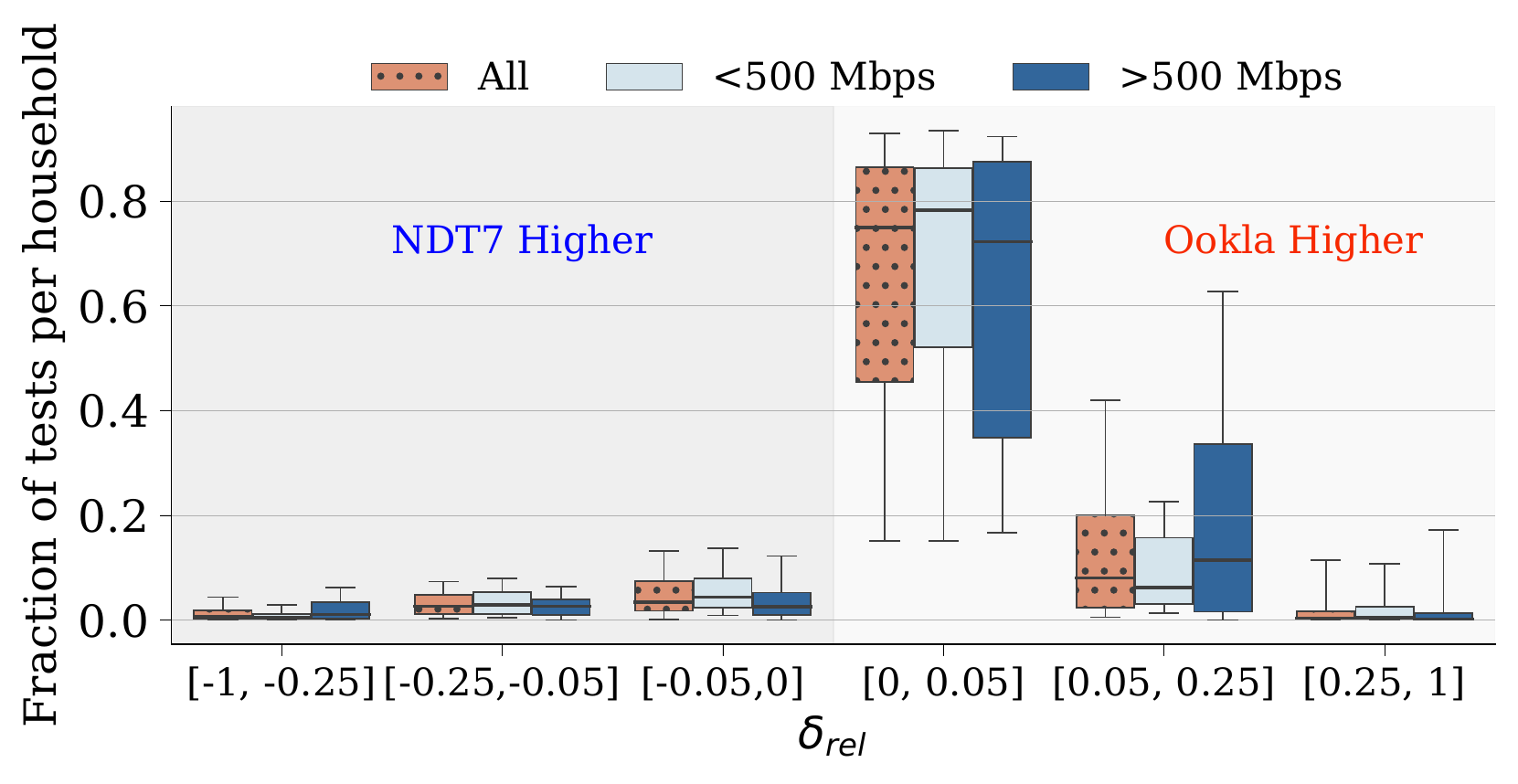}
	\caption{Download} \label{fig:down-diff} 
	\end{subfigure} \hfill
	\begin{subfigure}[t]{0.48\textwidth} 
	\centering 
	\captionsetup{width=.5\linewidth}
	\includegraphics[width=1\textwidth,keepaspectratio]{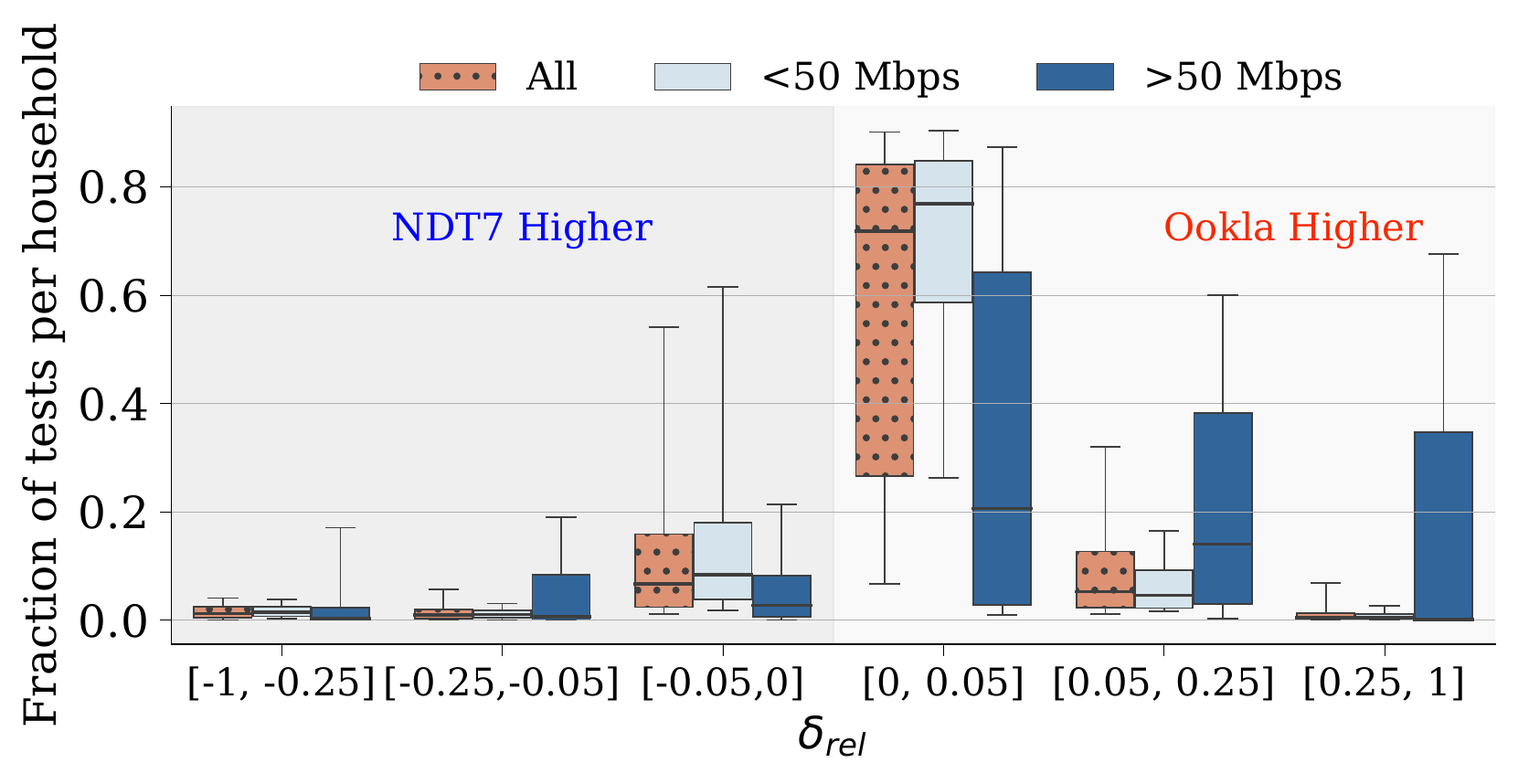}
	  \caption{Upload} \label{fig:up-diff} 
	  \end{subfigure}
	\caption{The distribution of each paired test class, partitioned by speed
	tier. Each point in a box plot is the fraction of tests that fall into that
	class for a given household. The boxes are first and third quartiles, the line inside the box is 
	median, whiskers are 10th and 90th percentile.}
	\label{fig:diff}
\end{figure*}

\subsubsection{How much do individual test results differ?}\hfill\\
\paragraph{Method.} We now study the difference in reported speeds for
each paired test. For our analysis, we first compute the signed relative
difference ($\delta_{rel}$) for each paired test:

\begin{equation}\label{eq:rel-diff}
	\delta_{rel} = \frac{S_{\ookla} - S_{\ndt}}{max(S_{\ookla}, S_{\ndt})}
\end{equation}

$S_{\ookla}$ and $S_{\ndt}$ are the speeds reported by \ookla and \ndt,
respectively. Dividing the difference by the maximum of two speed values bounds
$\delta_{rel}$ to $[-1, 1]$.  

\paragraph{Interpreting $\delta_{rel}$.} If $\delta_{rel} > 0$,
\ookla reported a higher speed than \ndt for that paired test, and  vice-versa
if $\delta_{rel} < 0$. The magnitude of $\delta_{rel}$ indicates the size of the 
difference between the two tests. For example, if $\delta_{rel} = 0.05$, then
the speed reported by \ndt is 5\% lower than the speed reported by \ookla.
Similarly, if $\delta_{rel} = -0.25$, then the speed reported by \ookla is 25\%
lower than the speed reported by \ndt.

For each paired test, we compute $\delta_{rel}$  and classify it
into one of six classes based on $\delta_{rel}$. Each class is defined by the
direction of the difference (whether \ookla or \ndt reports a higher speed) and
the magnitude of the difference. Table~\ref{tab:delta-rel-explainer} defines
the classes and how to interpret them. 

\begin{table}[h]
\small
	\centering
	\begin{tabular}{clr}
	\hline
	& Difference                      & Bin                                                   \\ \hline
																							   & Low                            & $0<\delta_{rel}\leq0.05$                              \\
																							   & \cellcolor[HTML]{EFEFEF}Medium & \cellcolor[HTML]{EFEFEF}$0.05<\delta_{rel}\leq0.25$   \\
	\multirow{-3}{*}{\begin{tabular}[c]{@{}c@{}}\ookla reports\\  a higher speed\end{tabular}} & High                           & $0.25<\delta_{rel}\leq1.0$                            \\ \hline
																							   & Low                            & $-0.05\leq\delta_{rel}<0$                             \\
																							   & \cellcolor[HTML]{EFEFEF}Medium & \cellcolor[HTML]{EFEFEF}$-0.25\leq\delta_{rel}<-0.05$ \\
	\multirow{-3}{*}{\begin{tabular}[c]{@{}c@{}}\ndt reports\\ a higher speed\end{tabular}}    & High                           & $-1.0\leq\delta_{rel}<-0.25$                          \\ \hline
	\end{tabular}
	\caption{Summary of paired test classes and how to interpret them.}
	\label{tab:delta-rel-explainer}
	\vspace{-2em}
\end{table}

We then compute the fraction of paired tests that fall into each category for
each household. This approach allows us to characterize the frequency of paired
tests with low, medium, and high difference in reported speed. We analyze the
distribution for each household instead of looking at the distribution across
all households for two reasons: 1) each household is unique in that the network
conditions present vary between households and 2) the number of paired tests
from each household is not uniform. Figure~\ref{fig:diff} shows the
distribution for each paired test class, where each point 
is the fraction of tests that fall into that class for a given household.

\paragraph{Results.} 
Figure~\ref{fig:down-diff}) shows that, in the case of downstream throughput, for most paired 
tests, \ookla
reports a slightly higher speed than \ndt. Across all households, the median
fraction of paired tests for which $0<\delta_{rel}\leq0.05$ is 91.3\%. It is
likely that differences of this magnitude are caused by differences in the test
protocol, for which \ookla's sampling mechanism discards low-throughput samples
and \ndt's does not. The fraction of paired tests with medium and high
difference is much lower. For medium differences, the median fraction of paired
tests for which $0.05<\delta_{rel}\leq0.25$ (\ookla reports a higher speed) is
9.6\%, while the median fraction of paired tests for which
$-0.25\leq\delta_{rel}<-0.05$ (\ndt reports a higher speed) is 3.9\%. As for
high differences, the median fraction of paired tests within [0.25, 1] (\ookla
reports a higher speed) is only 0.3\%, while the median fraction of paired tests
within [-1, -0.25] (\ndt reports a higher speed) is 0.6\%. It is thus
more common for a given household to have \ookla report a speed that is 25\%
lower than \ndt as opposed to vice versa. This finding aligns
with trends observed for the sample household shown in Figure~\ref{fig:similar},
where \ookla, while typically reporting slightly higher speeds, would
occasionally report a significantly lower speed.

We further partition households into \textit{high} and \textit{low} speed tiers.
A household is classified in the high downlink speed tier if the nominal
download speed (defined in Equation~\ref{eq:normalized-thruput}) is greater than
500 Mbps, otherwise it is classified in the low downlink speed tier. As for
uplink speed tiers, a household is classified as high-speed if the nominal
upload speed is greater than 50Mbps and low speed otherwise. As such, a
household may be in the high downlink speed tier but the low uplink speed tier.
We see that households in the high downlink speed tier (>500 Mbps) have a
greater fraction of paired tests with medium and high differences than
households in the low speed tier ($\leq $500Mbps). For example, there is a
median of 13.7\% of paired tests for which $0.05<\delta_{rel}\leq0.25$ among
high downlink speed tier households, while the corresponding fraction for low
speed downlink households is only 7.2\%. The trends are similar for paired
upload tests (see Figure~\ref{fig:up-diff}): the 75th percentile fraction of
paired tests within [0.25, 1] among high speed (> 50 Mbps) households is 34.7\%
compared to just 1.2\% for low speed (< 50 Mbps) networks. 

We believe these variations could result from two factors.
First, as the available throughput in real-world networks is variable, more so at high speed, the
flooding and sampling mechanisms have an impact on the reported speed. Our
in-lab experiments indicate that \ookla has clear advantages over \ndt in this
respect, given its use of multiple TCP connections, adaptive test length, and sampling 
strategy of discarding low throughput samples. Second, for high speed tiers, it is
possible that the bottleneck link for one or both the tests is not the ISP
access link but a link further upstream. 
\vspace{0.5em}
\begin{mdframed}[roundcorner=5pt, backgroundcolor=black!10] 
	\textbf{Takeaway}: The average difference in reported speed is either not significant or 
	within 5\% of the average test values in up to 73\% and 78\% households for download and 
	upload tests, respectively.  For individual test pairs, we observe significant differences -- 	
	the median fraction of paired tests for which \ndt reports a speed that is 0--5\% lower than 
	\ookla is 91.3\% and a speed that is 5--25\% lower is 9.6\%.
\end{mdframed}

\subsection{Effect of Server Selection}\label{subsec:server-selection}

\begin{table}[t]
\small
	\centering
	\begin{tabular}{@{}lllrr@{}}
	\toprule
						   &                           & \multicolumn{1}{l}{}                                      &                               &                                                                                 \\
	\multirow{-2}{*}{Rank} & \multirow{-2}{*}{AS Name} & 
	\multicolumn{1}{l}{\multirow{-2}{*}{AS Type}}             & \multirow{-2}{*}{\% of Tests} & 
	\multirow{-2}{*}{\begin{tabular}[l]{@{}r@{}}\% Households\end{tabular}} 
	\\ \midrule
	1                      & Nitel                     & ISP
                           & 24.4\%                        & 98.3\%                                                                         \\
	\rowcolor[HTML]{EFEFEF} 
	2                      & Puregig                   & ISP
                           & 13.8\%                        & 96.6\%                                                                          \\
	3                      & Whitesky                  & ISP
                           & 8.6\%                        & 87.3\%                                                                          \\
	\rowcolor[HTML]{EFEFEF} 
	4                      & Comcast                   & ISP
                           & 8.0\%                         &
                           80.6\%                                                                          \\
	5                      & Windstream                & ISP
                           & 6.9\%                         &
                           92.4\%                                                                          \\
	\rowcolor[HTML]{EFEFEF} 
	6                      & Frontier                  & ISP
                           & 5.9\%                         &
                           89.0\%                                                                          \\
	7                      & Cable One                 & ISP
                           & 5.7\%                         &
                           94.9\%                                                                          \\
	\rowcolor[HTML]{EFEFEF} 
	8                      & Rural Telecom             & ISP
                           & 4.1\%                         &
                           57.9\%                                                                          \\
	9                      & Hivelocity                &  Cloud Service& 3.0\%                         & 
	39.4\%                                                                          \\
	\rowcolor[HTML]{EFEFEF} 
	10                     & Enzu                      & Cloud Service & 2.7\%                         & 
	42.8\%                                                                          \\ \bottomrule
	\end{tabular}
	\caption{Top 10 Ookla Servers ranked by the number of tests that use that
	server. Percent of households indicates the fraction of households for which
	at least one test used the given server.}
	\label{tab:ookla-server}
\end{table}

\begin{table}[t]
\small
	\centering
	\begin{tabular}{@{}llr@{}}
	\toprule
						   &                           & \multicolumn{1}{c}{}                                                                              \\
	\multirow{-2}{*}{Rank} & \multirow{-2}{*}{AS Name} & \multicolumn{1}{c}{\multirow{-2}{*}{\begin{tabular}[c]{@{}c@{}}\% of Total\\ Tests\end{tabular}}} \\ \midrule
	1                      & GTT                       &
    19.62\%                                                                                           \\
	\rowcolor[HTML]{EFEFEF} 
	2                      & Tata Communications       &
    19.49\%                                                                                           \\
	3                      & Cogent                    &
    19.38\%                                                                                           \\
	\rowcolor[HTML]{EFEFEF} 
	4                      & Level3                    &
    19.36\%                                                                                           \\
	5                      & Zayo Bandwidth            &
    19.24\%                                                                                           \\ \bottomrule
	\end{tabular}
	\caption{NDT Servers ranked by the percentage of total tests}
	\label{tab:ndt-server}
\end{table}

\paragraph{Server infrastructure.} We first characterize the set of servers
observed over the course of our deployment. Across all households and
tests, we observe 32 unique \ndt servers. The top 15 most used \ndt
servers service 97.1\% of all \ndt tests. In light of this, we limit our
analysis to these 15 servers. Looking next at each \ndt server hostname, we find
five different Tier-1 networks, with 3 of the 15 servers co-located with each Tier-1
network (Table~\ref{tab:ndt-server}). 

As for \ookla, we observe far greater diversity in where servers are placed.
We observe 56 unique \ookla servers.
For the remaining analysis, we characterize the top 10 most commonly used \ookla servers. 
The top 10 servers account for 83.6\% of all \ookla tests conducted by 
our deployment fleet
(Table~\ref{tab:ookla-server}). The hostname, which we further validate with
an IP lookup, indicates that 8 of the 10 servers are co-located with consumer or
enterprise ISP network, while the remaining two are
co-located with a cloud provider. 

We next study how often households use each server. For \ndt tests, we find each
server is used roughly equally across tests. This is expected as \ndt uses
M-Lab's Naming Server to find the server~\cite{ndt-server-selection}. The default policy of 
the naming server
returns a set of nearby servers based on client's IP geolocation. The \ndt
client then randomly selects a server from this set.
As for \ookla, some servers are used disproportionately more than others. This
is in line with \ookla's server selection policy. The \ookla client pings a
subset of nearby test servers (selected using client IP geolocation) and picks
the server with minimum ping latency~\cite{ookla-server-selection}. 

\begin{figure}[t] 
\centering
\begin{subfigure}{.48\textwidth} \centering
\includegraphics[width=1\textwidth,keepaspectratio]{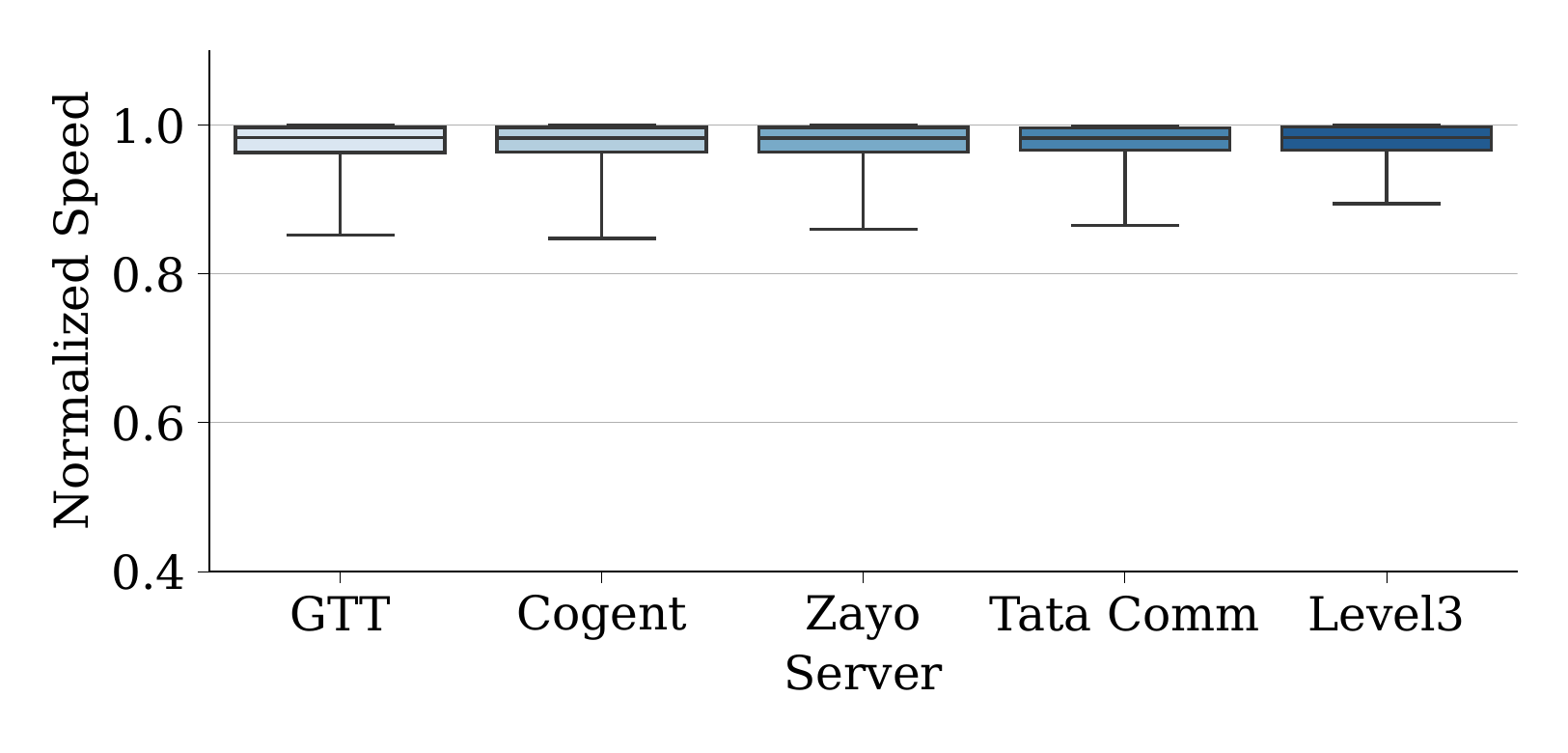}
\caption{\ndt} \label{fig:server-down-ndt7} 
\end{subfigure}%
 \hfill
\begin{subfigure}{.48\textwidth} 
\centering 
  \includegraphics[width=1\textwidth,keepaspectratio]{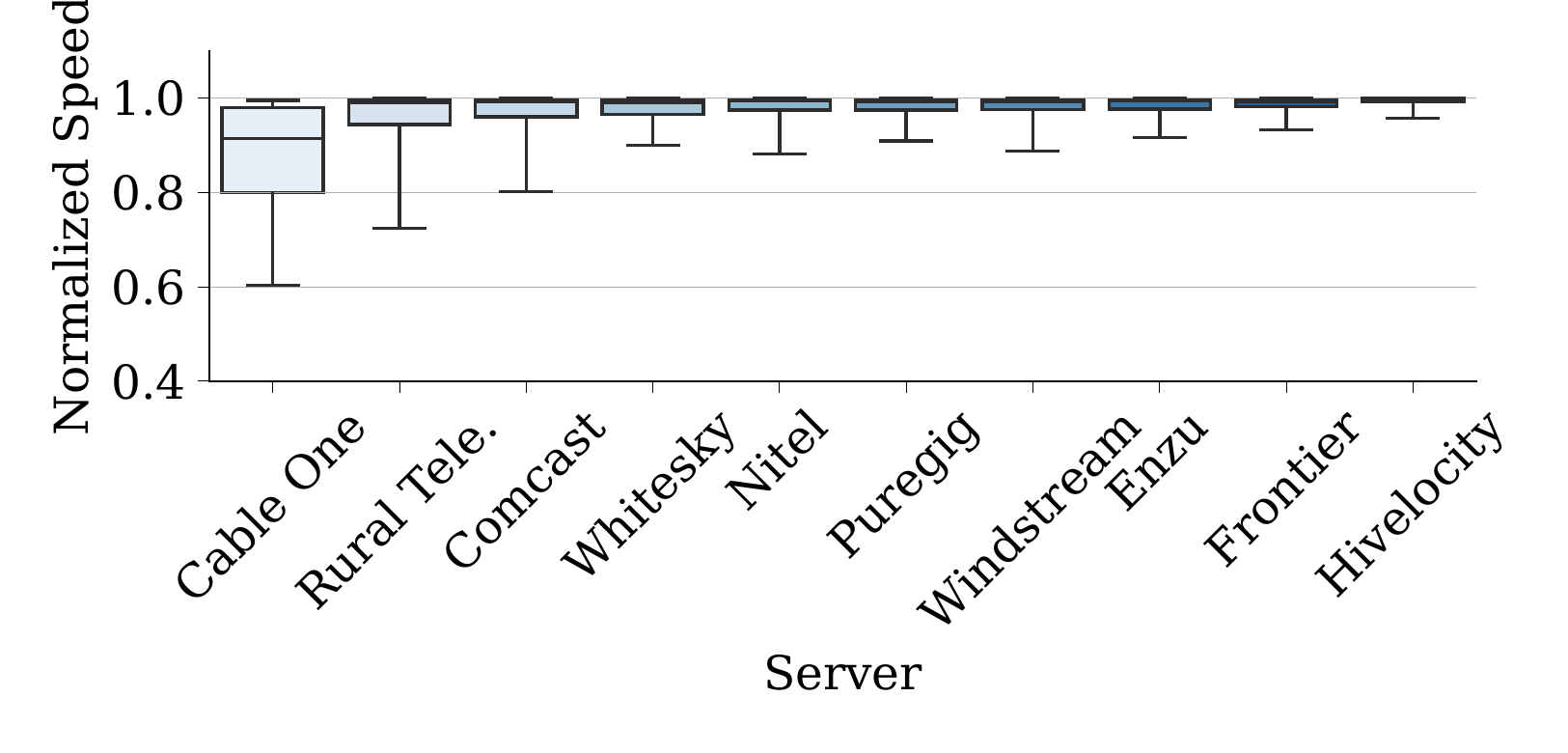}
  \caption{\ookla} \label{fig:server-down-ookla} \end{subfigure}
\caption{Distribution of normalized download speeds across servers. Note that the y-axis 
begins at 0.4.}
\label{fig:server-down} 
\end{figure}

\begin{figure}[t] 
	\begin{subfigure}{.48\textwidth} \centering
		\includegraphics[width=1\textwidth,keepaspectratio]{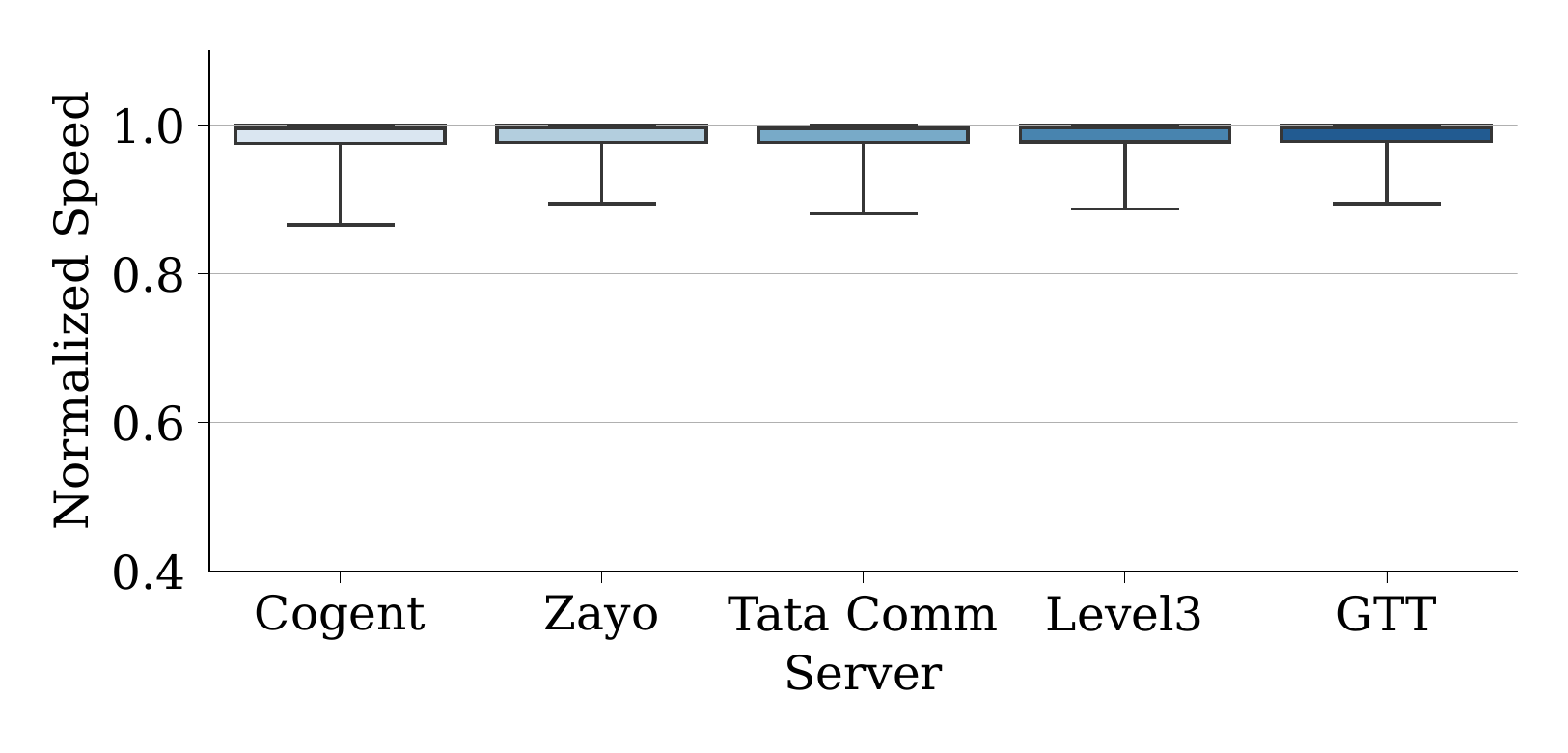}
		\caption{\ndt} 
		\label{fig:server-up-ndt7} 
	\end{subfigure} %
\hfill
	\begin{subfigure}{.48\textwidth} 
		\centering 
		\includegraphics[width=1\textwidth,keepaspectratio]{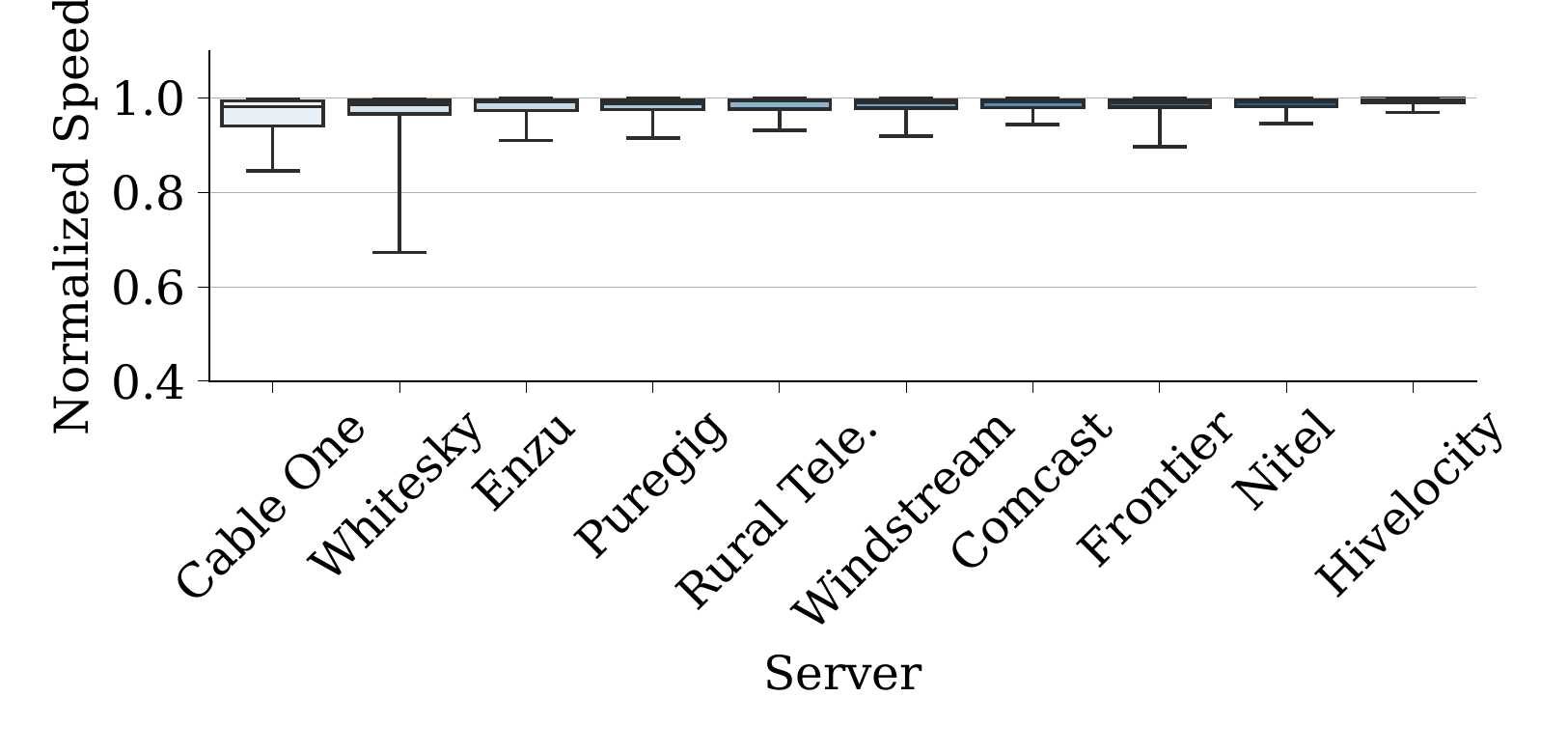}
		\caption{\ookla} 
		\label{fig:server-up-ookla} \end{subfigure}
	\caption{Distribution of normalized upload speed across servers. Note that the y-axis 
	begins 
		at 0.4.}
	\label{fig:server-up} 
\end{figure}

\begin{figure}[t]
	\centering
	\begin{minipage}[t]{0.48\textwidth}
	\centering
\includegraphics[width=\textwidth,keepaspectratio]{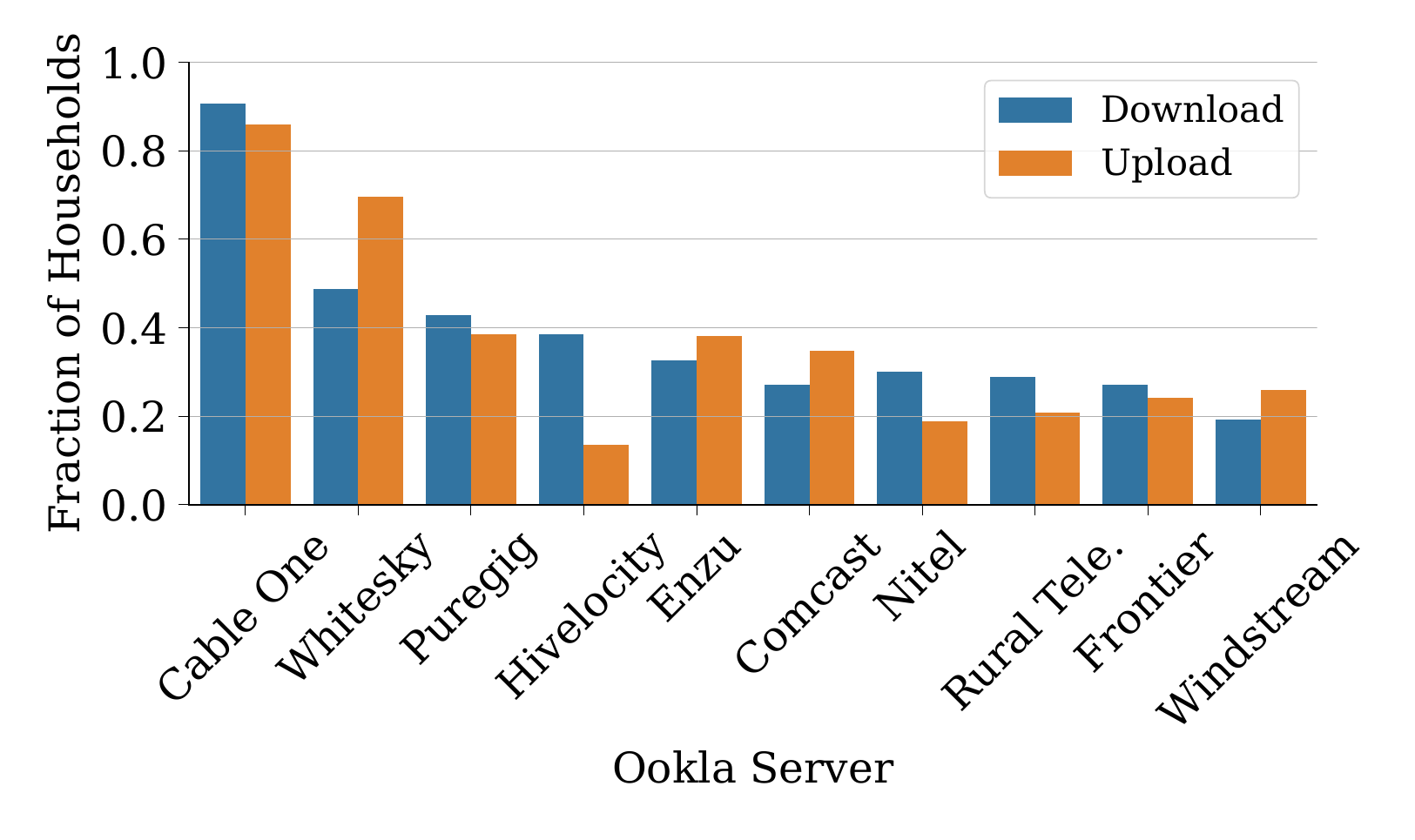}
\caption{Percentage of households for which a server is in the bottom
	three servers,
	ranked based on median normalized reported speed from tests using that server.}
\label{fig:ookla-server-ranking} 
	\end{minipage}%
 \hfill
	\begin{minipage}[t]{.48\textwidth}
	\centering
\includegraphics[width=\textwidth,keepaspectratio]{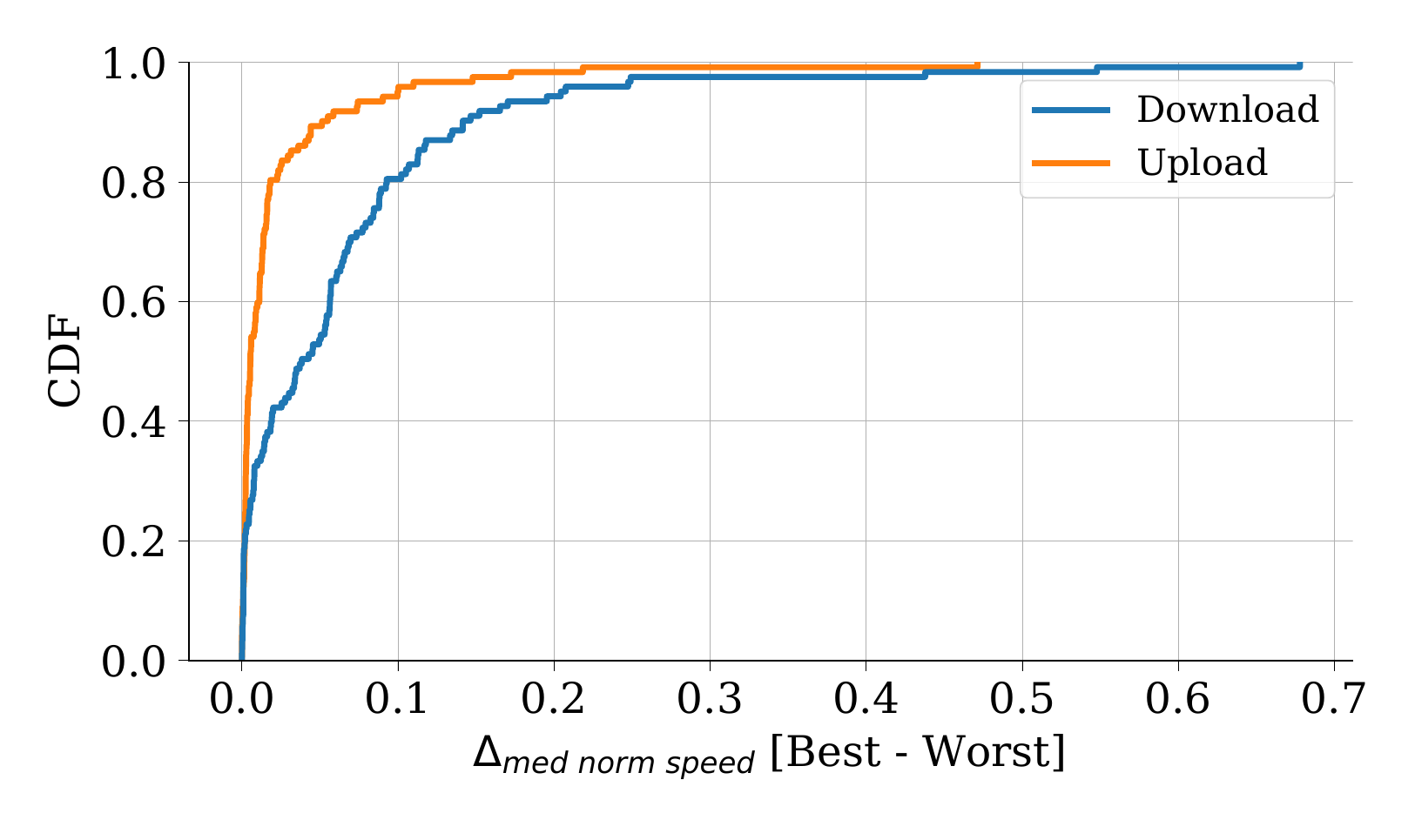}
\caption{Distribution of difference in median normalized speed between the best and worst 
	\ookla server across households. } 
\label{fig:server-ranking-best-worst} 
\end{minipage}
\end{figure}

\paragraph{Performance variation across servers.}
We now study how the reported speed varies with the choice of the
server. We group all test results by server and then compute the
normalized speed for each test, as defined in Equation~\ref{eq:normalized-thruput}. 
Figure~\ref{fig:server-down} shows the distribution of normalized download
speeds across different servers for both \ookla and \ndt. For \ndt, we find that
the distribution of normalized speed is similar across test servers. The bottom 10\% of tests 
across all \ndt servers have speeds at least 11\%-16\% lower than the household's nominal 
speed 
depending on the server. On the other hand, several  \ookla servers show higher variability 
than others. For instance, 
the median normalized speed for tests conducted using Cable One's
\ookla server is 0.91, suggesting that when using Cable One's server, 50\% of tests
report speeds at least 9\% lower than the household's nominal speed. In contrast, the median 
normalized speed for Rural Telecom's
\ookla server is 0.99.

For upload tests (Figure~\ref{fig:server-up}), we observe a similar trend. 
The distribution of normalized speed is similar
across \ndt servers, while several \ookla servers show greater variability than others. 
Specifically, for \ookla, Comcast's and Cable One's server report a median normalized speed 
of 0.99 and 0.98, respectively, while the $10^{th}$ percentile normalized speed
is 0.94 and 0.84, 
respectively. This heterogeneity across \ookla servers could be caused by either server 
underprovisioning or
issues in the end-to-end network path between the client and \ookla server. 

Because the number of tests per households varies, certain households with many
tests may be overrepresented in the overall trends. To determine whether these
trends hold for individual households, for each household, we rank 
\ookla servers by the median normalized speed from tests that used that server.
We remove servers for which fewer than 10 tests used that server. 
Figure~\ref{fig:ookla-server-ranking} shows the proportion of households for
which a given server is among the three lowest ranked servers for that household. 
Cable One's server is a bottom three server for 90\% and 85\% of
households, for download and upload tests, respectively. Similarly, Whitesky's server is a 
bottom three server for 48\% and
69\% households, for download and upload tests, respectively. 

Finally, we quantify the difference between \ookla servers within a household by computing 
the median normalized speed of the highest and the
lowest ranked server, denoted by $\Delta_{med\ norm\ speed}$. 
Figure~\ref{fig:server-ranking-best-worst} shows the CDF of
$\Delta_{med\ norm\ speed}$ across households. For download tests, the
worst servers report a difference of 0.1 or greater for 20\% of the users. The
difference in upload tests is smaller, with 5\% users reporting a
difference of at least 0.1, likely because most access links have slower upload
than download speeds.

\vspace{0.5em}
\begin{mdframed}[roundcorner=5pt, backgroundcolor=black!10] 
	\textbf{Takeaway}: The majority of \ndt
	servers reside in transit ISPs, whereas \ookla's open server participation
	policy leads to some of their servers being placed in consumer ISPs and
	cloud providers. Although Ookla's server policy may lead to a greater
	coverage, there may be issues of quality assurance; we observe that several
	\ookla servers systematically under-reporting speed.
\end{mdframed}

\subsection{Effect of Time of Day}\label{subsec:time-of-day}
Past work has suggested that ISPs experience higher loads at certain ``peak''
usage periods, leading to a corresponding increase in network 
congestion~\cite{sundaresan2011broadband}. If
true, the time of day at which a speed test is run may affect the reported
speed. In this section, we analyze our deployment data to see if the effect is similar for \ookla 
and \ndt. In accordance with previous work~\cite{sundaresan2011broadband}, we define peak 
hours as 7--11~p.m. 

For this analysis, we use a \textit{two sample
t-test}, or independent t-test, because we are comparing two independent
populations. We define our null hypothesis $H_{0}$ to be that for a given
household, the average reported speeds for tests conducted in peak and off-peak
hours are equal. To satisfy the normality condition, we only test 
households
for which we have at least $30$ tests from both peak and off-peak hours.
Enforcing this condition leaves $91$ households with sufficient download tests
and $93$ households with sufficient upload tests. Recall that we
treat households that undergo speed tier upgrades as two separate households,
explaining why the number of households differ for download and upload tests. As in 
Section~\ref{subsec:t-test-paired}, we use a significance level ($\alpha$) of 0.01.

\begin{table}[t]
\footnotesize
	\centering
	\begin{tabular}{@{}clrr@{}}
	\toprule
	\multicolumn{1}{l}{Direction} & Tool & \multicolumn{1}{l}{\# Rejecting $H_{0}$} & 
	\multicolumn{1}{l}{\% Rejectng $H_{0}$}  \\ \midrule 
								  & \cellcolor[HTML]{EFEFEF}\ndt &
	\cellcolor[HTML]{EFEFEF}37    & 
								  \cellcolor[HTML]{EFEFEF}40.7\% \\
	\multirow{-2}{*}{\begin{tabular}[c]{@{}c@{}}Download\\ (n = 91)\end{tabular}} & 
	\ookla                       & 21
								  & 23.1\%                         \\ \midrule &
	\cellcolor[HTML]{EFEFEF}\ndt & \cellcolor[HTML]{EFEFEF}7     &
	\cellcolor[HTML]{EFEFEF}7.5\% \\
	\multirow{-2}{*}{\begin{tabular}[c]{@{}c@{}}Upload\\ (n = 93)\end{tabular}}
								 & \ookla                       & 9 & 9.7\%
							 \\ \bottomrule \end{tabular}
	\caption{Results of t-test where $H_{0}$ is that the average reported speeds for
	tests conducted in peak and off-peak hours are equal.} 
	\label{tab:tod-test}
\end{table}

Table~\ref{tab:tod-test} summarizes the results.
For download tests, we find sufficient evidence to reject $H_{0}$
for \ndt on 40.7\% of households, but only 23.1\% of networks for \ookla. This
difference suggests that the time of day at which the speed test is conducted has
a greater effect for \ndt than for \ookla. Furthermore, the set of households
for which we can reject $H_{0}$ for \ndt is a strict superset of the set of
households for which we can reject $H_{0}$ for \ookla. For this fact and the
fact that we run paired tests, we can attribute the discrepancy in the number of
networks to the specific tool, not the access network (which we control
for). The different behavior of \ookla and \ndt can be due to one or 
both of the following: (1)~the client-server paths for \ndt have more cross-traffic during peak 
hours than \ookla; (2)~the test protocol, especially use of multiple threads and sampling 
heuristic by \ookla, makes \ookla more resilient to cross traffic than \ndt. 
For upload tests, we can only reject $H_{0}$ on 7.5\% and 9.7\% of
households for \ndt and \ookla, respectively; this decrease may be the result
of asymmetry in upload and download traffic loads.

\section{Related Work}\label{sec:related}

\paragraph{Speed test design.} There are two primary
ways to measure throughput: (1) packet probing and (2)
flooding. Most packet probing techniques send a series of packets and infer metrics like 
available bandwidth or link capacity based on the inter-arrival packet 
delay~\cite{keshav1991control, dovrolis2001packet, ribeiro2003pathchirp, jain2002pathload, 
hu2003evaluation}. 
More recently, Ahmed et al.~\cite{ahmed2020flowtrace} estimate bandwidth
bottlenecks by probing the network using recursive in-band packet trains. 
However, these techniques can be inaccurate especially for high speed networks due to their 
sensitivity to packet loss, queuing policy etc.
As a result, most commercial speed tests, including ones offered by both
ISPs~\cite{att2022speedtest, comcast2022speedtest} and non-ISP
entities~\cite{ookla2022speedtest, mlab2022speedtest, netflix2022fast}, are
flooding-based tools that work by saturating the bottleneck link through active measurements. 

\paragraph{Evaluating speed tests.} Feamster and
Livingood~\cite{feamster2020measuring} discuss considerations with using
flooding-based tools to measure speed. They do not, however, conduct empirical
experiments to characterize \ndt and \ookla performance. Similarly, Bauer et
al.~\cite{bauer2010understanding} explain how differences in speed test design
and execution contribute to differences in test results. Bauer et al.'s work
differs from ours in several ways. First, both \ookla and NDT have seen major
design changes in the 12 years since that study. Both tools have updated their
flooding and sampling mechanisms, and NDT's latest version (\ndt) uses TCP BBR
instead of TCP Reno. Second, they only analyze public NDT data and do not study
both \ookla and NDT in controlled lab settings, nor did they conduct paired
measurements in the wide area that allows direct comparison of \ookla and NDT,
as we do. Complimentary to our comparative analysis is work by Clark et
al.~\cite{clark_measurement_2021} that provides recommendations on how to use
aggregated NDT data, including considering the self-selection bias and other
end-user bottlenecks like slow WiFi and outdated modems.

\paragraph{Residential broadband.} Goga et al.~\cite{goga2012speed} evaluate
the accuracy of various speed test tools in residential networks, yet tools
have changed and speeds on residential networks have increased more than
$20\times$ since this study ten years ago. Sundaresan et
al.~\cite{sundaresan2011broadband} studied network access link performance in
residential networks more than ten years ago. Whereas our work is more focused
on characterizing speed test tools, this work examined network performance
differences across ISPs, looking at latency, packet loss, and jitter in
addition to throughput. Canadi et al.~\cite{canadi2012revisiting} use publicly
available \ookla data to analyze broadband performance in $35$ metropolitan
regions. Finally, the Federal Communications Commission (FCC) conducts the
Measuring Broadband America project (MBA)~\cite{fcc2022mba}, an ongoing study
of fixed broadband performance in the United States. The FCC uses SamKnows
whiteboxes~\cite{samknows2022} to collect a suite of network QoS metrics,
including throughput, latency, and packet loss.  Because the MBA project maps
broadband Internet performance across different ISPs, they use a single speed
test---a proprietary test developed by SamKnows---and do not consider \ookla
or \ndt. Most recently, Paul et al.~\cite{paul2022context} analyze
crowdsourced data from M-Lab and Ookla speed tests.  That work focuses on
contextualizing speed test data that has already been collected by inferring
the end-user broadband subscription plan from existing measurements.  This
paper studies a different, complementary concern: how the speed test
methodology affects the reported speeds.  Although
they also find that \ookla reports higher median speeds than \ndt, our 
approach is quite different. We compare back-to-back \ndt and \ookla
tests from individual households, while Paul et al. compare aggregate test
data from the same ISP and same inferred subscription plan at a city-level. 

\balance\section{Conclusion}
\label{sec:discussion} 

This paper provided an in-depth comparison of \ookla and \ndt, focusing on both
test design and infrastructure. Our measurements, both under controlled network
conditions and wide area network using paired tests, present new insights about
the differences in behavior of the two tools. Yet, while this work is the {\em
first} to perform such a controlled and extensive comparison between these speed
test tools, we neither hope nor expect that it will be the last, and note that
many important questions remain, both in the technical and policy realms.  Below
we summarize the implications of our findings for the future of speed test tools
and data analysis and outline multiple avenues for future work. 

\paragraph{\ndt and \ookla, not \ndt or \ookla.}
Some readers may (mis)interpret our results as an endorsement or condemnation
of a particular measurement tool---but that is not our intention at all. We
believe, based on our findings---and the notion that there are {\em many}
facets to Internet speed---that the existing datasets can all be useful,
provided that the users of the data understand the measurement techniques used
to gather the data, as well as the limitations of each tool that may make it
more (or less) appropriate for specific questions. Both \ndt and \ookla
provide a wealth of measurements, and each dataset offers valuable
measurements from the edge of the Internet, as well as complementary coverage
and perspective. \ndt in particular also offers an open-source tool and method
that can be rigorously tested (and improved upon); indeed, this very research
has allowed us to highlight bugs and shortcomings of the \ndt tool and
infrastructure that the developers have since fixed. It is worth noting,
however, that the tools are
designed, implemented, and deployed differently, and therefore measure
slightly different phenomena, resulting in divergent results under different
network conditions and circumstances. While no single approach is ``correct'' or
``incorrect'',  it is critical to understand how different network
conditions and deployments---from cross-traffic to high latency to
interconnect congestion---may affect the numbers that each of these tools may
report, as well as how their results may diverge.

It is also important to note that the design of speed tests---and NDT in
particular---has evolved over time, and thus warrants continual re-appraisal.
For instance, past work found that NDT is limited by its use of a single TCP
connection, leading to it underestimating speed on high capacity
paths~\cite{feamster2020measuring, bauer2010understanding}. Our results,
however, indicate that for most in-lab and real-world cases, \ookla and \ndt
report similar speeds. \ndt's improved results are likely due to many factors,
including improvements to the tool implementation, to browsers and operating
systems as well as migration to different transport protocols (e.g., TCP BBR).
As such, our work provides a methodology to comprehensively assess these tools
as they continue to evolve. Nevertheless, we recommend greater transparency into 
the test design and updates to enable practitioners to better understand the
reported speed.

\paragraph{Implications for speed data analysis.} Our study reveals specific
network conditions
where \ookla and \ndt yield inaccurate results:
\begin{itemize} 
	\item \ndt reports lower speeds when the client-server latency
	is high (e.g., RTT > 100ms for a 100~Mbps link) and data from such high
	latency tests should be discarded. \item The client software version can
	impact the test accuracy. For instance, we found the  the initial version of
	\ndt client we used was over-reporting upload speeds as it relied on
	client-side measurements. \ndt software releases are accompanied with
	release notes that can help identify if (and how) the older version impacted
	test accuracy. 

	\item  The longitudinal WAN measurements highlight
	significant performance differences between \ookla's servers. This warrants
	the need for careful auditing of server-side issues and even server
	selection methods that currently only rely on latency measurements. An
	alternative could be to use results from past measurements of clients within
	the same IP subnet to select optimal server. \end{itemize}

More generally, it is important to account for the network conditions, testing context 
(e.g., device and software version), and test parameters (e.g., test duration, number of 
TCP connections) while analyzing the speed data. We recommend that the speed test 
tools 
provide such metadata along with speed results for a more accurate analysis of speed 
data.

\paragraph{How generally do these results apply?} Our in-lab experiments were
conducted over a comprehensive set of network conditions, and we thus expect
the results from the in-lab experiments to apply broadly. We conducted these
controlled in-lab experiments to insulate the tools from the effects of a wide range of
network conditions, from consumer devices to WiFi effects, that can occur in
real-world deployments and skew crowdsourced measurements. The controlled
experiments vary only one network parameter at a time to isolate the effect of
that parameter on the accuracy of these widely used speed tests. To our
knowledge, a controlled, comparative study between these tools has never been
performed at this scale---and certainly not for the modern versions of \ndt
and \ookla. Readers and users of these tools should be able to read the
results from the in-lab experiments to gain a general understanding of how
these tools are likely to behave under certain circumstances. 

Moreover, the in-lab results can be extended to the design and use of other
speed tests that rely on flooding the bottleneck link to measure throughput.
This work highlights the effects of a number of design choices, including the
test length, number of TCP connections, and sampling technique on the speed
reported by each test.  These characteristics are not specific to the tests
that we study in this paper, and designers and users of of these and others
speed tests (and their data) can draw insights from the findings in this
study. For example, the results of Section~\ref{sec:in-lab} confirm
that a speed test tool that uses multiple TCP connections or runs for a
longer duration will report higher speeds under high latency and background
traffic than a tool that uses a single TCP connection or runs for a shorter
duration. 

The wide-area experiments have a different purpose: To understand how these
tools might produce divergent measurements in operational access networks. To
do so, we performed a longitudinal study for nine months, across nearly 126
homes in a large metropolitan area, covering about 30 neighborhoods and all of
the major ISPs for that city. Although the sample is comprehensive and
longitudinal, it is important to recognize that this sample does omit certain
ISPs, and in particular certain {\em types} of ISPs. For example, the sample
does not contain fixed 5G providers, which is an increasingly common mode of
home Internet access in certain geographies. Our deployment study, however, is
ongoing, and the measurement tools and analysis that we have produced is
open-source and public. Several other regions in the country are now, in fact,
adopting our software and tools for their own studies. To this point,
conducting a wide-area study of these tools in different geographies and
settings, with a different sample, continues to be an excellent avenue for
ongoing and future work---particularly as both Internet access and the tools
to measure it continue to evolve.

\section{Acknowledgments}
This work was supported by National Science Foundation awards CNS-2224687 and
CNS-2223610 and a {\tt data.org} Inclusive Growth and Recovery Challenge Award.
We thank the reviewers and our shepherd, Zubair Shafiq, for helpful comments. We
are also grateful to Guilherme Martins, Marc Richardson, and Grace Chu for their
help in building the data collection platform and recruiting participants for
the study of residential networks.

\end{sloppypar}
\bibliographystyle{ACM-Reference-Format}
\bibliography{paper}

\pagebreak
\label{lastpage}\clearpage
\appendix
\thispagestyle{empty}

\setcounter{figure}{0}
\renewcommand\thefigure{B.\arabic{figure}}

\setcounter{table}{0}
\renewcommand\thetable{A.\arabic{table}}

\section{Statement of Ethics}
We obtained approvals from our Institution Review Board (IRB) before deploying RasPi  
devices
in participants' households. We have uploaded the consent form that was shared with 
each participants before signing up. At each step in the study, we have taken utmost 
care about user privacy. The RasPi devices deployed in the household 
collect only active measurements. In fact, we can not monitor any user network traffic 
due to our network setup. We also remove any privacy-sensitive user identifiers (e.g., 
physical address, demographics) before analyzing the data. Taking a broader view of 
ethics, our work has the  potential for positive impact on society. It informs appropriate 
use of speedtest data which in turn has implications on accurately mapping and bridging 
the digital divide. 

\section{In-Lab}

\subsection{Network Conditions}
\subsubsection{Bandwidth}

\begin{figure}[H] 
	\centering
	\includegraphics[width=0.5\textwidth,keepaspectratio]{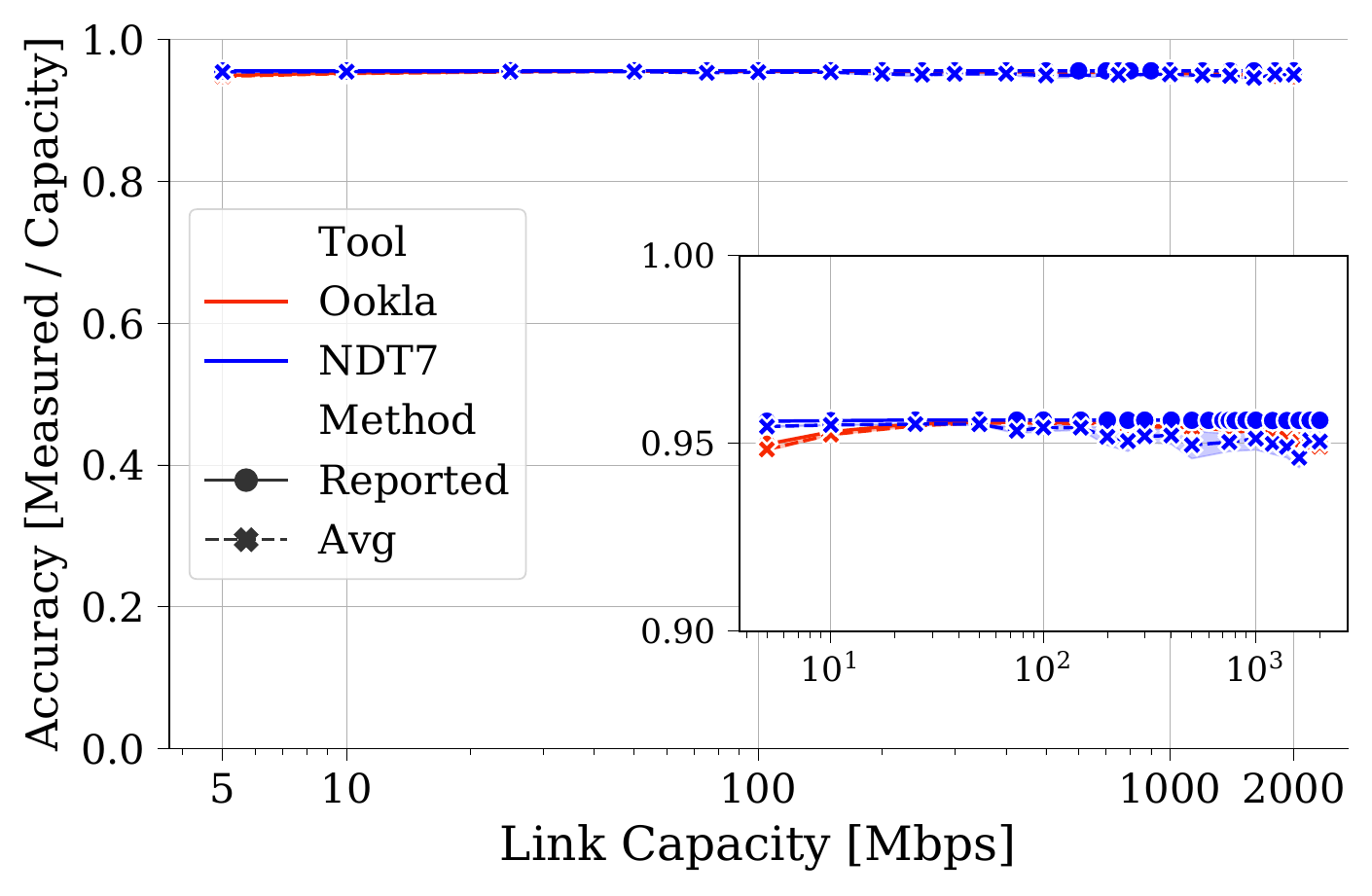}
	\caption{Tool performance vs. link capacity for upload tests. Note that
	the y-axis begins at 0.9. Shaded region represents a 95\% confidence
interval for $n=10$ tests.} 
	\label{fig:rate-upld} 
\end{figure}

\subsubsection{Latency}
\begin{figure}[H] 
	\centering
	\includegraphics[width=0.5\textwidth,keepaspectratio]{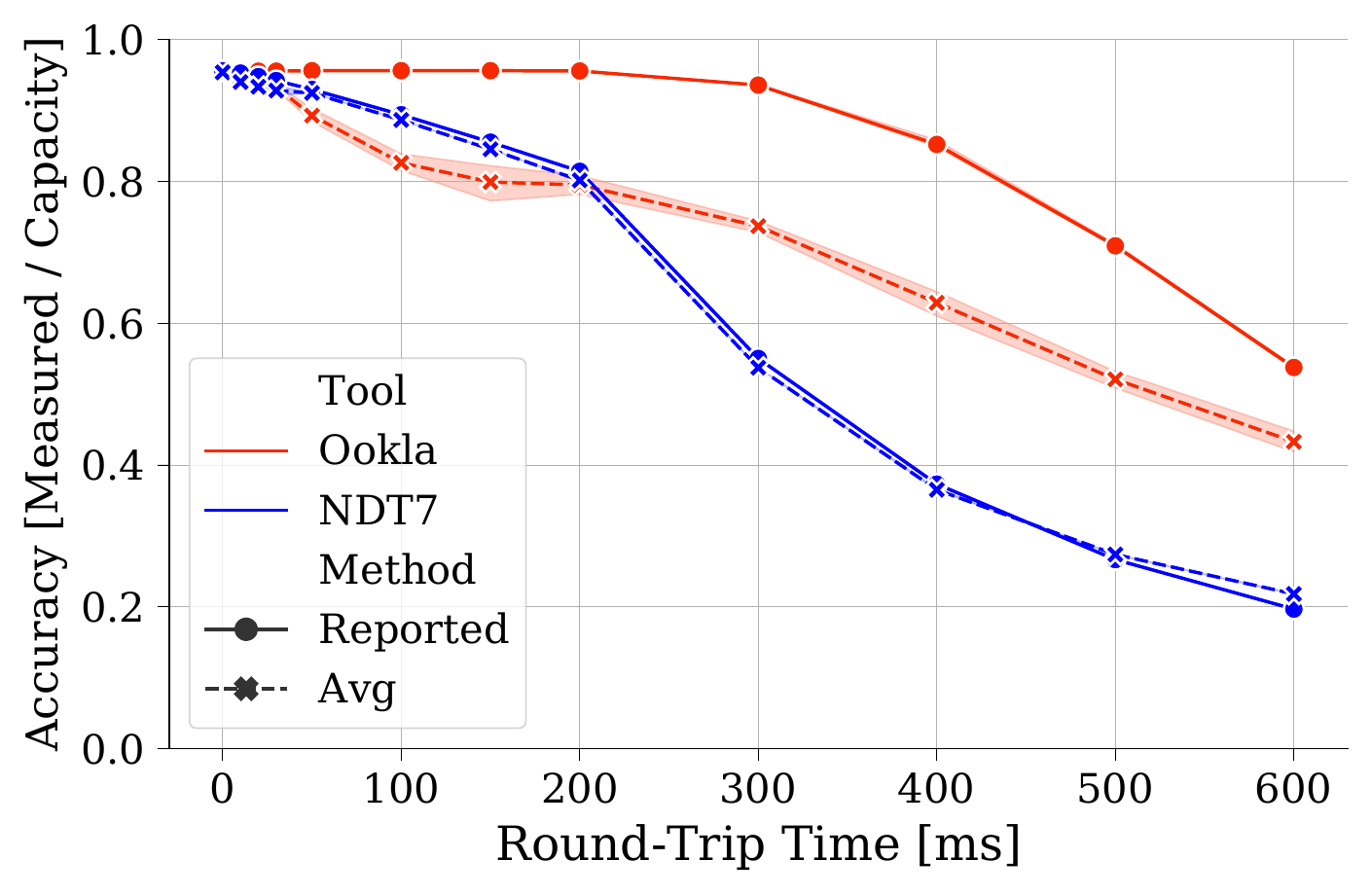}
	\caption{Tool performance vs. one-way latency for upload tests. Shaded
	region represents a 95\% confidence interval for $n=10$ tests.} 
	\label{fig:latency-upld} 
\end{figure}

\begin{figure}[H] 
	\centering
	\includegraphics[width=0.5\textwidth,keepaspectratio]{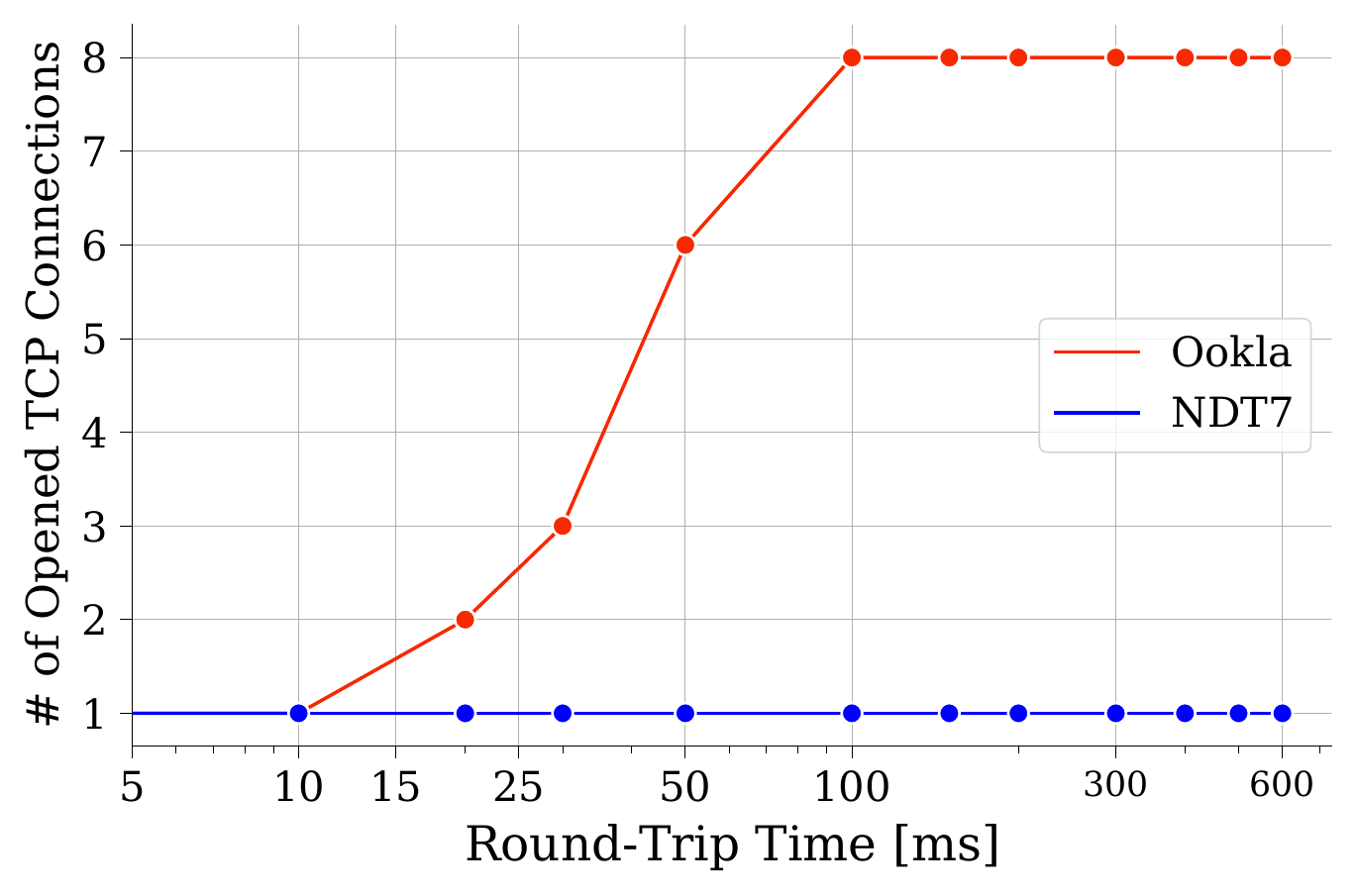}
	\caption{Number of TCP connections opened during a speed test vs. one-way
	latency between the server and the client.} 
	\label{fig:latency-nthreads} 
\end{figure}

\begin{figure}[H] 
	\centering
	\includegraphics[width=0.5\textwidth,keepaspectratio]{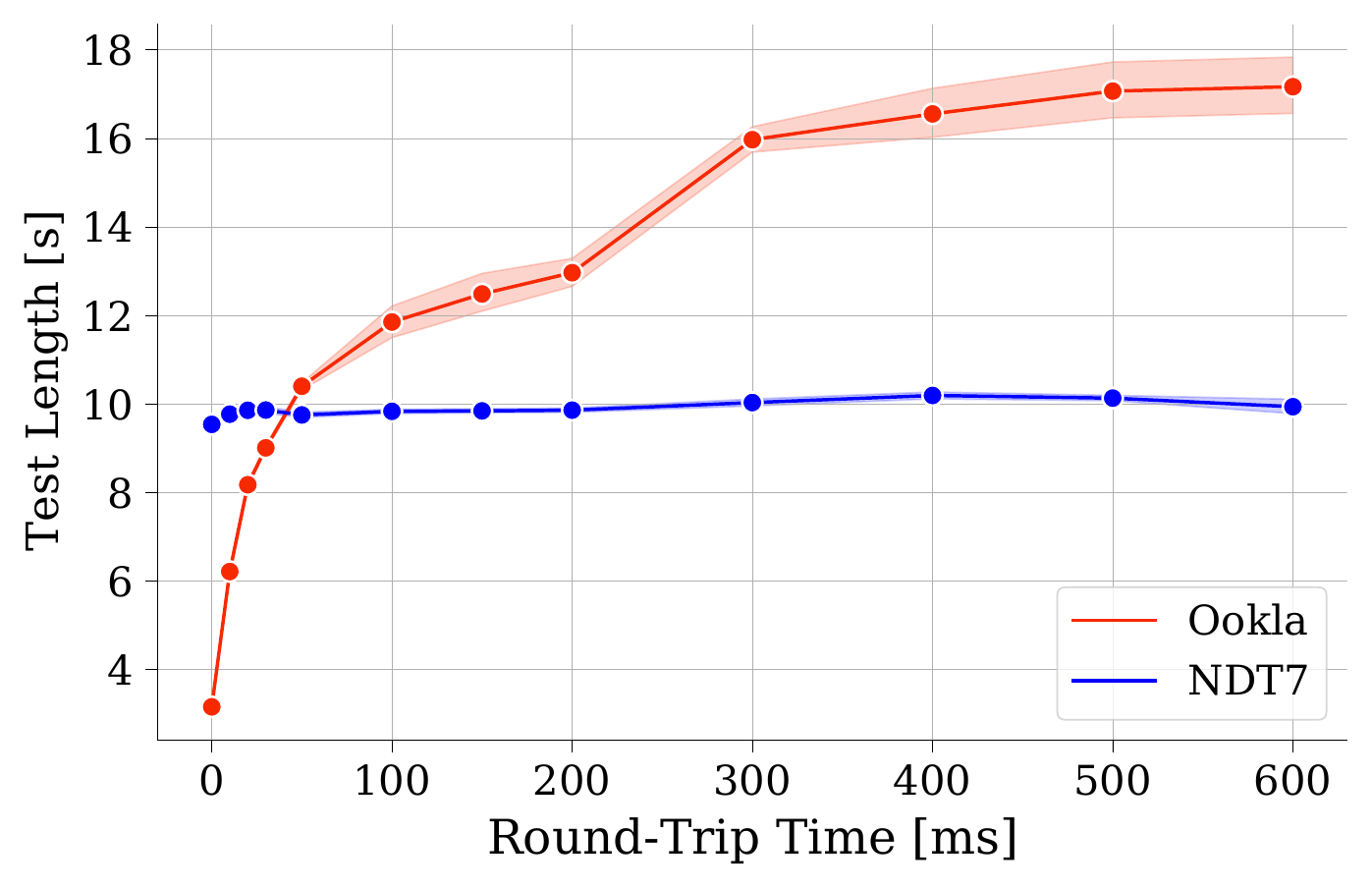}
	\caption{Length of a speed test as one-way latency increases.} 
	\label{fig:latency-testlen} 
\end{figure}

\subsubsection{Loss}
\begin{figure}[H] 
	\centering
	\includegraphics[width=0.5\textwidth,keepaspectratio]{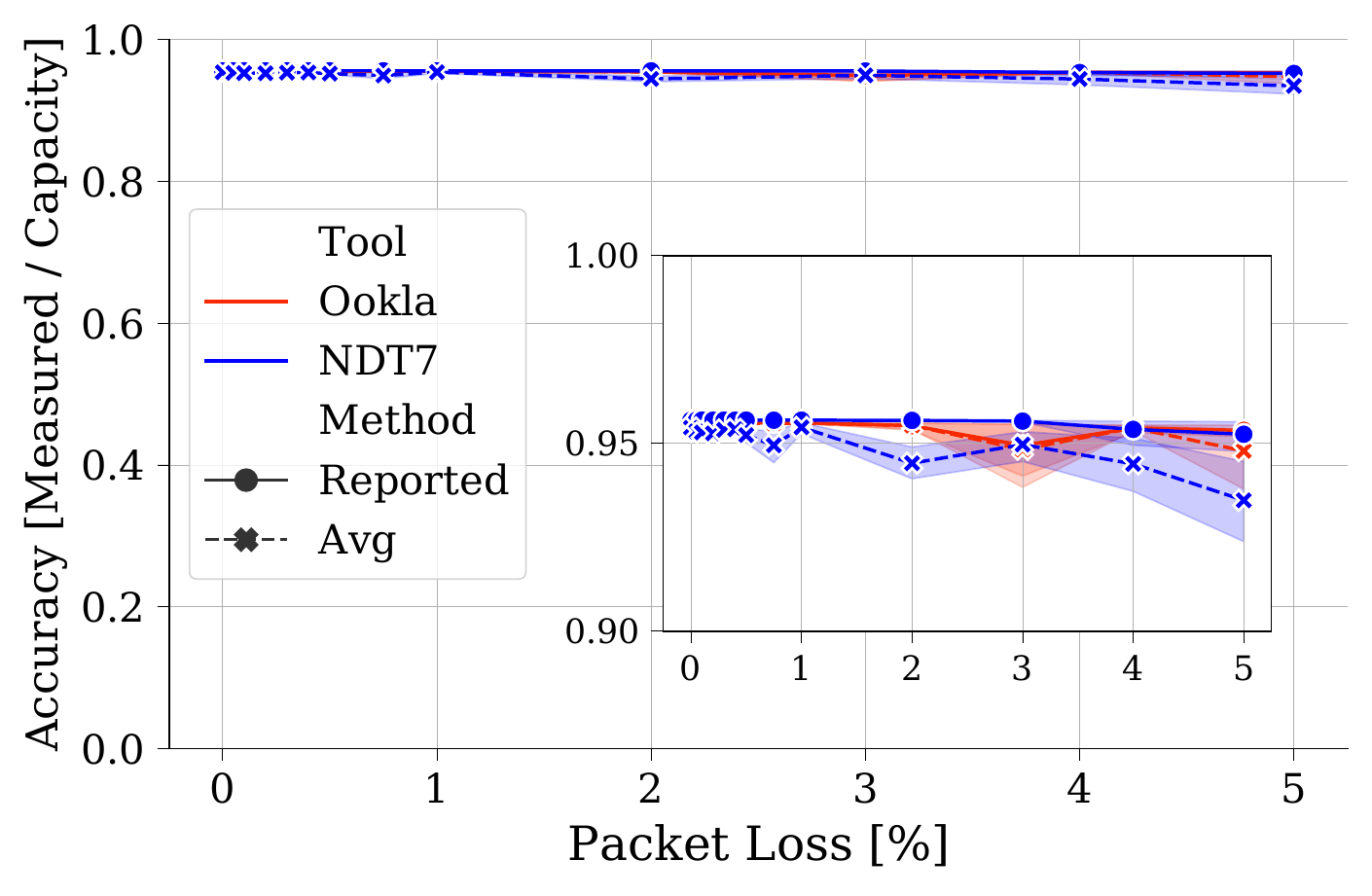}
	\caption{Tool performance vs. packet loss for upload tests. Shaded
	region represents a 95\% confidence interval for $n=10$ tests.} 
	\label{fig:loss-upld} 
\end{figure}

\subsubsection{Client Type}

\begin{figure*}[h!]
    \begin{subfigure}[h]{0.33\textwidth}
    		\centering
        \includegraphics[width=\textwidth,keepaspectratio]{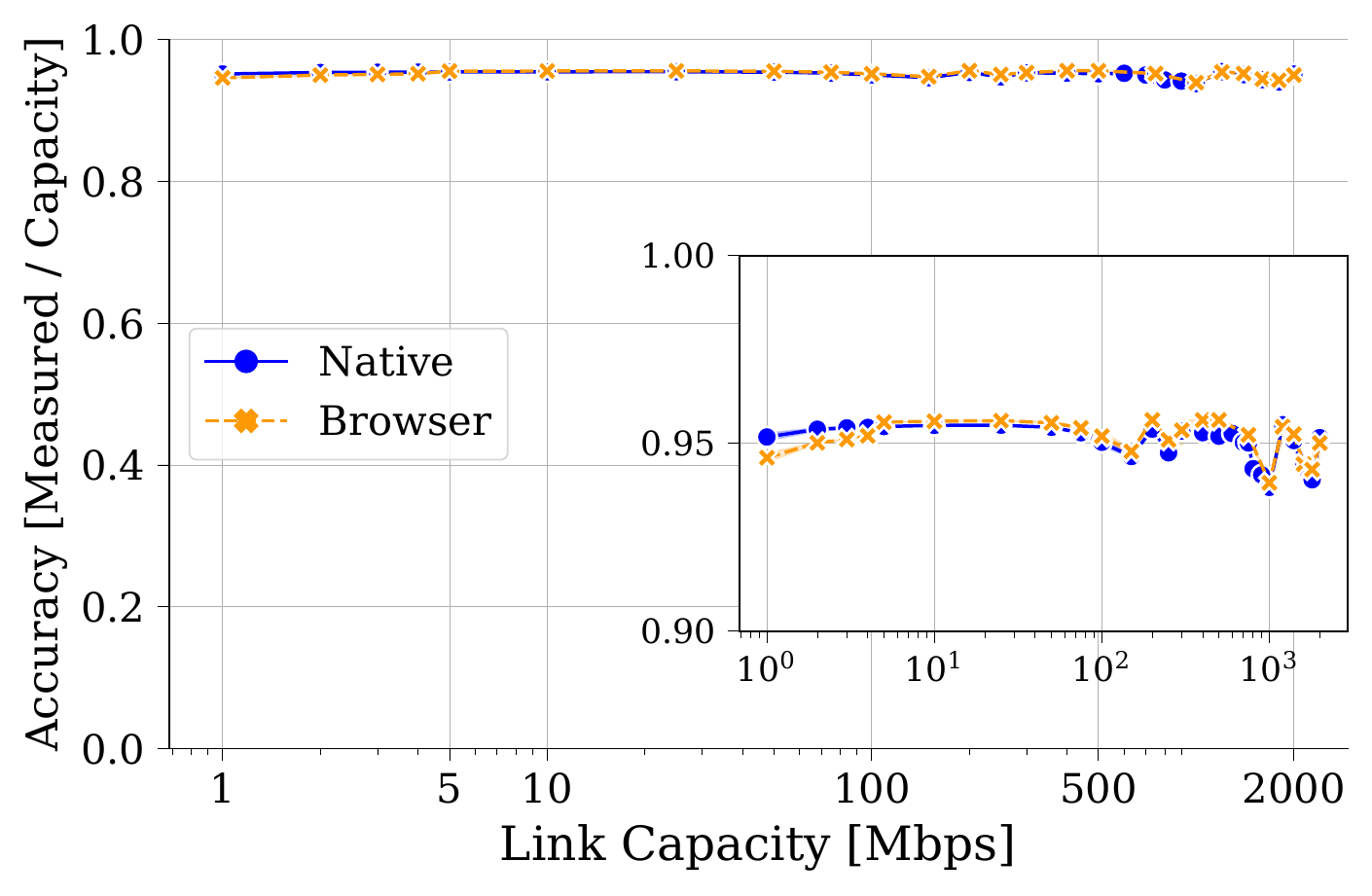}
        \caption{Download - Client Type vs. Link Capacity}
 		\label{subfig:browser-rate-down}
    \end{subfigure}%
    \hfill
	\begin{subfigure}[h]{0.33\textwidth}
        \centering
        \includegraphics[width=\textwidth]{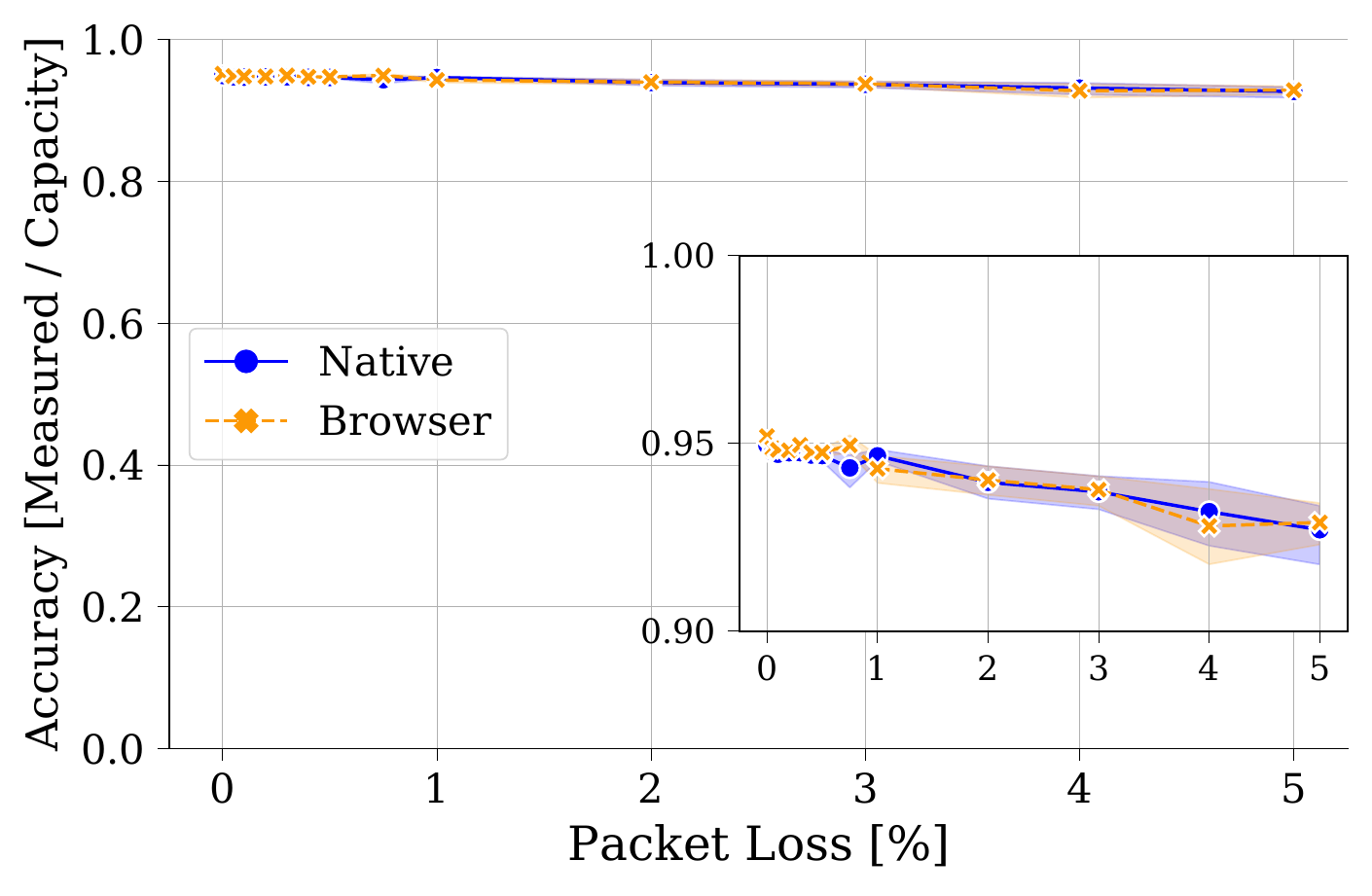}
    \caption{Download - Client Type vs. Packet Loss}
    \label{subfig:browser-loss-down}
    \end{subfigure}%
    \hfill
	\begin{subfigure}[h]{0.33\textwidth}
        \centering
        \includegraphics[width=\textwidth]{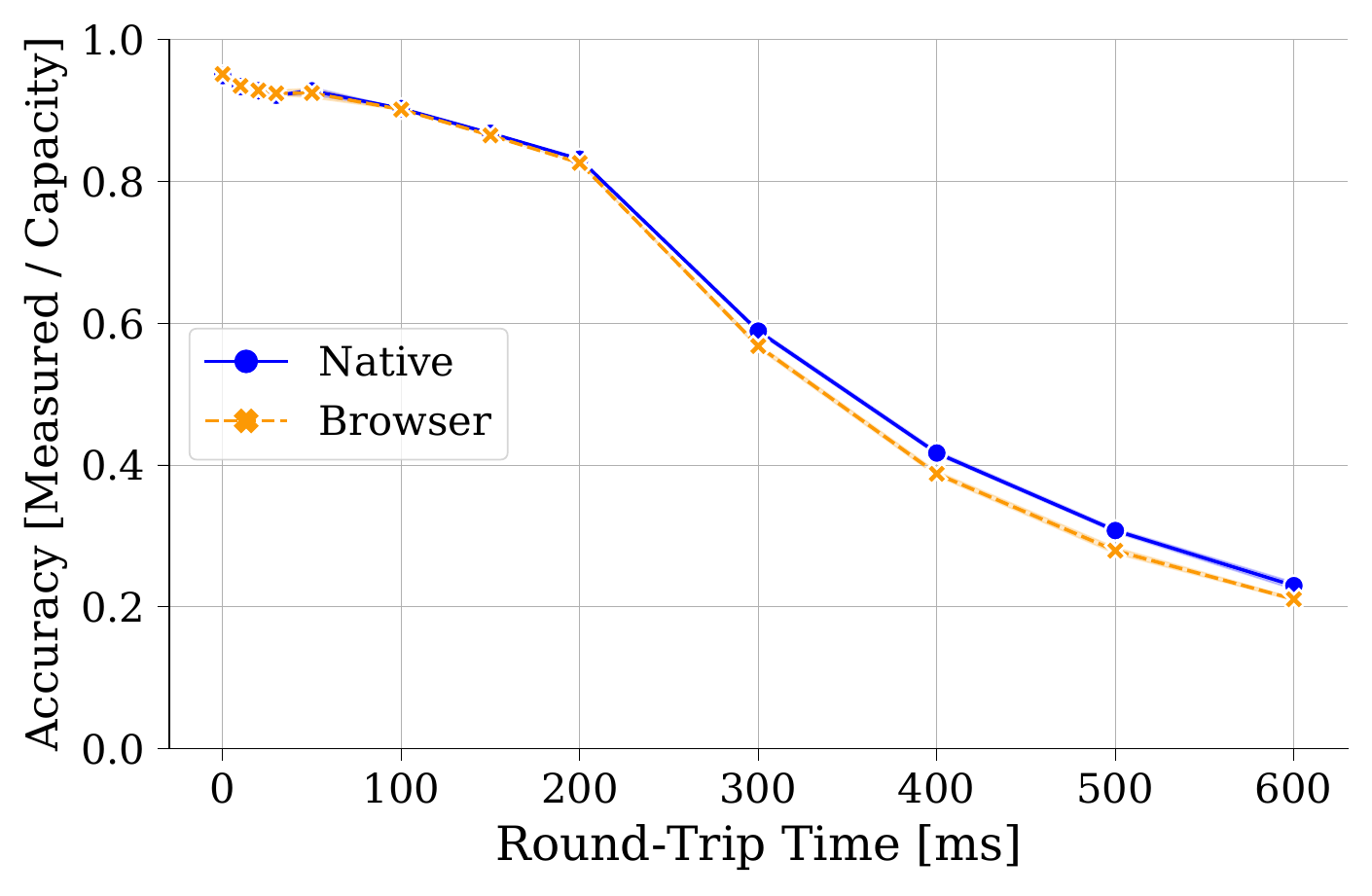}
    \caption{Download - Client Type vs. Latency}
    \label{subfig:browser-latency-up1}
    \end{subfigure}
	\caption{Tool performance on different client types (browser vs. native) for
	download tests.}
    \vspace{-1em}
	\label{fig:browser-down}
\end{figure*}

\subsection{Real-World Deployment}

\begin{figure}[h]
    \begin{subfigure}[t]{0.48\textwidth}
    		\centering
        \includegraphics[width=\textwidth,keepaspectratio]{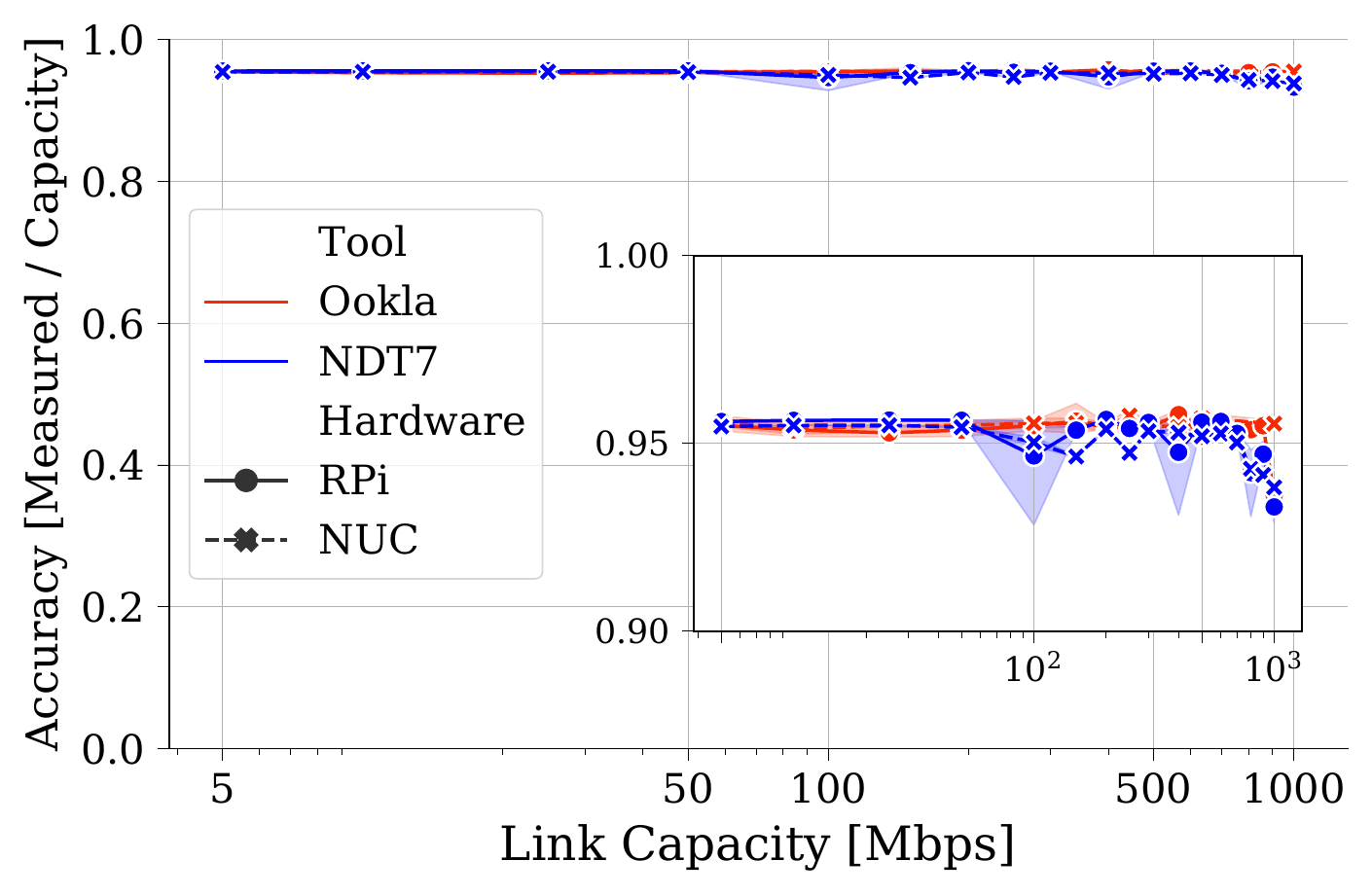}
        \caption{Download}
 		\label{subfig:hardware-rate-down}
    \end{subfigure}%
    \hfill
	\begin{subfigure}[t]{0.48\textwidth}
        \centering
        \includegraphics[width=\textwidth]{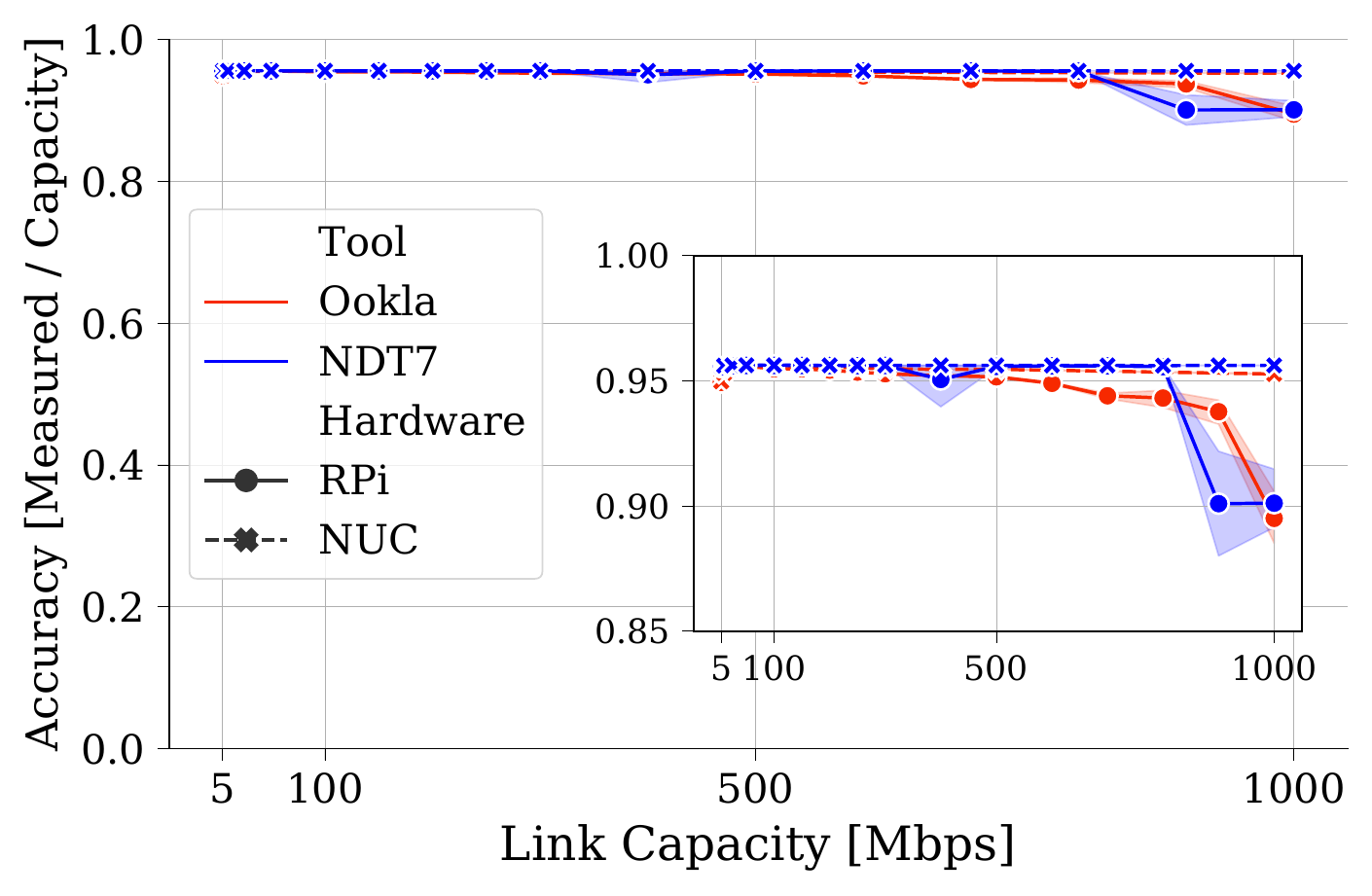}
    \caption{Upload}
    \label{subfig:hardware-rate-up}
    \end{subfigure}%
    \hfill
	\caption{Performance vs. Link Capacity for Raspberry Pis and NUCs}
    \vspace{-1em}
	\label{fig:hardware-rate}
\end{figure}

\begin{figure}[h] 
	\centering
	\includegraphics[width=0.5\textwidth,keepaspectratio]{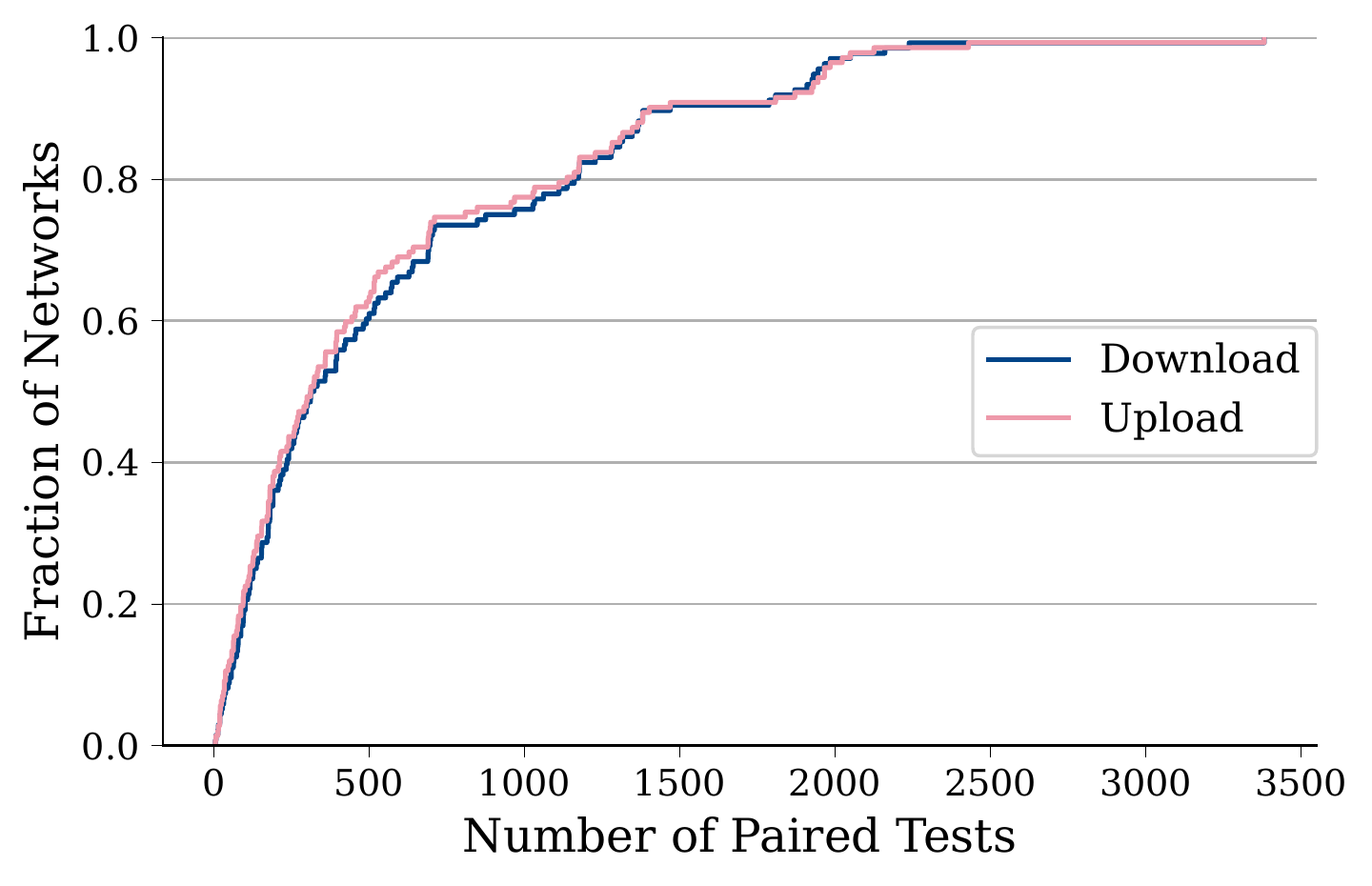}
	\caption{CDF of the number of paired tests across all tested networks} 
	\label{fig:ntests} 
\end{figure}

\end{document}